\documentclass[
reprint,
superscriptaddress,
amsmath,
amssymb,
prx,
]{revtex4-2}

\usepackage{graphicx}%
\usepackage{dcolumn}%
\usepackage{bm}%
\usepackage{hyperref}
\usepackage{xcolor}
\usepackage[english]{babel}
\usepackage[autostyle]{csquotes}
\usepackage[final]{changes}
\usepackage{cancel}

\begin{document}

\title{Light-Matter Interactions Beyond the Electric Dipole Approximation \\ in Extended Systems Without Multipole Expansion}

\author{Rishabh Dora}
\affiliation{
    Department of Chemistry, University of Rochester, Rochester, NY, USA}

\author{Roman Korol}
\affiliation{
    Department of Chemistry, University of Rochester, Rochester, NY, USA}
\affiliation{Current address: D\'epartement de Chimie, Universit\'e de Sherbrooke, Sherbrooke, Qu\'ebec J1K 2R1, Canada}

\author{Vishal Tiwari}
\affiliation{
    Department of Chemistry, University of Rochester, Rochester, NY, USA}
\affiliation{Current address: Department of Chemistry, University of Washington, Seattle, Washington 98195, USA}
\author{Rahul Chourasiya}
\affiliation{
    Department of Chemistry, University of Rochester, Rochester, NY, USA}

\author{Ignacio Franco}
\email{ignacio.franco@rochester.edu}
\affiliation{
    Department of Chemistry, University of Rochester, Rochester, NY, USA}
\affiliation{
    Department of Physics and Astronomy, University of Rochester, Rochester, NY, USA}
\affiliation{
    The Institute of Optics, University of Rochester, Rochester, NY, USA}

\date{\today}
\begin{abstract}
We present a general theoretical framework to capture light-matter interactions beyond the electric-dipole approximation, applicable to extended nano- and microscale materials interacting with spatially structured electric fields without truncation at finite multipolar order. The approach is based on the Power-Zienau-Woolley (PZW) Hamiltonian for light-matter interactions and a representation of the material’s Hamiltonian in a basis of maximally localized Wannier functions (MLWFs), which can be obtained from first-principles electronic structure calculations. We utilize this novel approach to clarify the limitations of the ubiquitous dipole approximation. To that end, we consider electric fields generated by light with both uniform and non-uniform intensity profiles and a range of ratios of system size to the wavelength of light. Through this analysis, we identify the precise conditions under which the dipole approximation breaks down, leading to significant errors in the light-induced dynamics. Contrary to conventional belief, we find that the dipole approximation is remarkably robust for uniformly illuminated 1-D or 2-D materials if light propagates in the direction perpendicular to the material. For 3-D materials or non-perpendicular illumination of lower-dimensional materials, conventional wisdom holds, and the dipole approximation starts to break down when the wavelength of light becomes comparable to the system size. Furthermore, the dipole approximation fails to capture the light-matter interaction when the material is illuminated partially or non-uniformly. For slowly varying field intensities this failure can be corrected by the finite-order multipolar corrections. However, for fields that vary substantially over the spatial extent of the system size, correcting via finite order multipolar terms becomes computationally impractical. In contrast, our approach captures the beyond-dipole light-matter interactions at a computational cost of a standard dipole calculation. This efficiency paves the way for accurate first-principles simulations of spatially-structured light-matter dynamics in nanoscale devices, quantum materials, and interfaces.
\end{abstract}
\keywords{Light-matter interactions, Beyond dipole approximation, PZW Hamiltonian }

\maketitle

\section{Introduction}

Light-matter interactions are foundational across physics, chemistry, materials science, biology, and emerging quantum technologies. They provide the spectroscopic means to probe and understand nature, and powerful handle to optically control the behavior of matter on ultrafast timescales~\cite{Mandel1995,CohenTannoudji1997,Brabec2000,Moerner2002,Cho2008,Haug2008,Krausz2009,Mukamel2009,Yu2010,Milonni2010,Shapiro2012,Craig2012,Kruchinin2018,Ghimire2018,Dombi2020,Torre2021,Goulielmakis2022,Heide2024a} .
Recent advances have yielded spectroscopies with unprecedented resolution~\cite{Xiao2017,Kraus2018,Biswas2022,Zhang2022,Kim2022,Zong2023,Boschini2024}, the ability to probe and control matter on femto to attosecond timescales \cite{Franco2007,Franco2008,Ghimire2010,Schultze2012,Higuchi2017,Ciappina2017,GarzonRamirez2020,Hui2021,Cavaletto2024,LHuillier2024}, 
the emergence of Floquet engineered materials~\cite{Oka2019,Shan2021,Zhou2023,Ito2023,Tiwari2025a,Tiwari2025b,Choi2025,Merboldt2025},
petahertz electronics~\cite{Schoetz2019,Boolakee2022,Hassan2024,Heide2024,Sennary2025,Feher2025}, and light-induced emergent phenomena~\cite{McIver2019,Zhang2024,UzanNarovlansky2024,Beaulieu2024,Zhang2025,Keren2026}.

A central approximation in the vast majority of theoretical and experimental studies is the electric-dipole approximation (EDA), in which the incident electric field is assumed to be spatially homogeneous across the sample and magnetic effects
are considered to be negligible \cite{CohenTannoudji1997,Mukamel2009,Yu2010,Schueler2021}. 
Formally, the expanded form of the spatial part of the electric field, $e^{i \mathbf{k} \cdot \mathbf{r}} = 1 + i \mathbf{k} \cdot \mathbf{r} - \frac{1}{2} (\mathbf{k} \cdot \mathbf{r})^2 + \dots $, can be truncated at the zeroth-order term in $\mathbf{k}$, where $\mathbf{k}$ is the wavevector with norm $k = |\mathbf{k}| = \frac{\omega}{c} = \frac{2 \pi}{\lambda} $, $\omega$ is the angular frequency, $\lambda$ the wavelength of the field, $c$ the speed of light, and $\mathbf{r}$ the position vector \cite{CohenTannoudji1997}. The validity of EDA is ensured when $ \mathbf{k} \cdot \mathbf{r} \ll 1$, due to the small spatial extent of the system relative to the electromagnetic wavelength (long-wavelength limit) \cite{Chernyak1995,CohenTannoudji1997,Mukamel2009}.

To go beyond the dipole approximation, it is customary to include finite-order multipolar corrections in the light-matter interaction Hamiltonian along with the dipole term. Such corrections have been investigated in the context of photo-ionization by X-rays \cite{Lindle1999,Bernadotte2012,Demekhin2014,List2015,Mohan2026} where the wavelength can be comparable to the molecular size, in two-photon ionization \cite{Hofbrucker2020} where electric quadrupolar interaction is required and in chiral spectroscopies probing optical activity where both electric and magnetic dipoles can play a role \cite{Condon1937,Ayuso2022,Horn2022}.
Non-dipole effects \added{can be deliberately engineered \cite{Ji2019,Ji2024}, and} have \added{also} been studied extensively in the context of strong-field light-matter interactions where the leading order magnetic field effects are considered \cite{Foerre2006,Bandrauk2006,Mishra2012,Ludwig2014,Brennecke2018,Jensen2020,Foerre2022,Jensen2022,Suster2023,DellaPicca2023,Jensen2025}. Although finite-order multipolar corrections can be used to extend EDA, the resulting interaction Hamiltonian depends on the choice of expansion point \cite{Bernadotte2012,Lestrange2015}. Beyond-dipole spectral response have been explored for molecular systems \cite{Bernadotte2012,List2015,Soerensen2019,Aurbakken2024} where minimal-coupling origin-independent approaches without truncation were developed. Recently, beyond-dipole effects within a QED framework have been explored where radiation back reaction was also taken into consideration for quantum systems \cite{Bonafe2025}. 
In parallel, the full multipolar Hamiltonian has been used to examine near-field interactions at the molecular length scale to describe photoexcitation of molecules with highly structured light emanating from nanoparticles \cite{Iwasa2009,Yamaguchi2015,Yamaguchi2016a,Noda2017,Noda2019,Iwasa2024,Nishizawa2025}. 

However, existing beyond-dipole frameworks are either limited to finite-order multipolar corrections that suffer from origin dependence \cite{Lestrange2015} and slow convergence in inhomogeneous fields \cite{Horn2023}, or to full minimal‑coupling and multipolar approaches whose computational cost has so far confined applications to small molecular systems \cite{List2015,Nishizawa2025} and model quantum materials \cite{Bonafe2025}. Here, we introduce a beyond-dipole framework for realistic extended systems interacting with spatially structured far-fields that avoids any multipolar truncation and its origin dependence, retains the numerical efficiency of standard dipole approximation calculations, and quantitatively resolves the interplay between spatial field inhomogeneity and the breakdown of the dipole approximation for nano- and micro-scale materials.

To account for the full spatial structure of an arbitrary electric field we use the semi-classical multipolar Power-\replaced{Zienau}{Zieneau}-Woolley (PZW) Hamiltonian \cite{Power1959,Woolley1971,Mukamel2009}. 
This Hamiltonian features a compact form for the light-matter interaction term. It is described as a spatial integral involving the dot product of the system's overall polarization and the applied electric field (see Eq.~\ref{PZW_Ham}). Although this Hamiltonian is well known and captures the full spatial dependence of the electric field, its use in atomistically detailed computations without invoking a multipolar expansion has remained an open challenge. Here, we develop a practical computational framework to perform quantum dynamics with the multipolar PZW Hamiltonian by using maximally localized Wannier functions (MLWFs) \cite{Marzari1997,Marzari2012} as a basis to describe the material Hamiltonian and its interaction with light. The MLWFs are highly localized and can be used to develop a diagonal representation of the position operator, greatly simplifying the computation of the matrix elements of the PZW Hamiltonian and of the multipolar expansions to all orders. This approach allows us to accurately capture the spatial structure of light-matter interactions in extended systems without a significant increase in computational cost. Further, our computational scheme can be used in conjunction with any electronic structure code as long as the resulting orbitals can be localized. Thus, the approach we propose enables investigation of light-matter interactions beyond the dipole approximation using atomistically detailed models.

To demonstrate the approach, we perform simulation in a simple 1-D model of \emph{trans}-polyacetylene (\emph{t}PA) chain. Further, to focus on the emerging physics and to probe the limits of the EDA we permit ourselves to vary the unit-cell length. Specifically, we investigate the effects of considering the full spatio-temporal profile of the field on the laser-induced dynamics, and assess the validity of the dipole approximation by varying system size with respect to illuminated area and the wavelength. To demonstrate the approach in a realistic setting, we further analyze the beyond-dipole effects using a nonuniform electric field generated via electrodynamic simulations by solving Maxwell's equations using the Finite-Difference Time-Domain (FDTD) method in a complex electromagnetic environment.
Contrasting the multipolar and the dipolar results, we can identify the physics that is emerging due to multipolar effects and quantify the limits of applicability of the EDA.
Our study not only advances the theoretical understanding of light-matter interactions, but also offers a practical computational tool for modeling these phenomena in realistic materials and devices.
 
The structure of this paper is as follows: section \ref{Hamiltonian} presents the theory of light-matter interaction beyond the dipole approximation; sections \ref{wannierbasis} and \ref{derivation} express the multipolar PZW Hamiltonian in terms of MLWFs; the details of equations of motion are given in Sec.~\ref{EqOfMotion}. Section \ref{model} gives the details of the model systems, the electrodynamic simulation of light and the quantum dynamics simulation parameters;
section \ref{results} presents and discusses the key findings. We summarize our work in Sec.~\ref{conclusion}.

\section{Theory} \label{theory}
\subsection{General Hamiltonian} \label{Hamiltonian}
Consider a material interacting with a laser with space $\mathbf{r}$ and time $t$ dependent \added{transverse component of} electric field $\mathbf{E}(\mathbf{r},t)$. The total Hamiltonian is $\hat{H} = \hat{H}_\mathrm{M} + \hat{H}_\mathrm{LM}(t)$, where $\hat{H}_{\mathrm{M}}$ describes the matter Hamiltonian and 
\begin{equation} \label{PZW_Ham}
    \hat{H}_{\mathrm{LM}} = - \int d^3 \mathbf{r} \ \hat{\mathbf{P}}(\mathbf{r}) \cdot \mathbf{E}(\mathbf{r},t) 
\end{equation}
is the \added{semi-classical} light-matter interaction; the much weaker magnetic-field interaction terms \cite{Mukamel2009} are not taken into account explicitly, but can also be added to this analysis. \added{The PZW Hamiltonian in Eq.~\ref{PZW_Ham} is equivalent to the standard velocity gauge representation of the minimal-coupling formulation. The two Hamiltonians are connected via a gauge transformation \cite{Babiker1983,Chernyak1995,CohenTannoudji1997,Mukamel2009,Woolley2020}.} The quantity $\hat{\mathbf{P}}(\mathbf{r}) = \sum_{j} \hat{\mathbf{P}}_{j} (\mathbf{r}) $ is the total polarization operator of the system of charges indexed by $j$.
The polarization of a single charge $q_j$ located at position $\hat{\mathbf{r}}_{j}$ is given by
\begin{equation} \label{polarization}
    \hat{\mathbf{P}}_{j}(\mathbf{r}) = \int_0^1 du\ q_{j} \hat{\mathbf{r}}_{j} \delta[\mathbf{r} - u \hat{\mathbf{r}}_{j}  ].
\end{equation} 
The integration in Eq.~\eqref{polarization} with respect to $u$ ensures the correct coefficients in the Taylor series expansion, leading to dipole, quadrupole, and higher-order multipole terms  \cite{CohenTannoudji1997, Mukamel2009}.  
We now follow the standard derivation of the multipolar expansion \cite{Mukamel2009}. The Dirac delta distribution $\delta[\mathbf{r} - u \hat{\mathbf{r}}_{j}]$ on the right hand side of Eq.~\eqref{polarization} can be expanded in a Taylor series with small displacement $u\hat{\mathbf{r}}_j$ about the origin $\mathbf{r}=\mathbf{0}$
\begin{equation} 
    \hat{\mathbf{P}}_{j}(\mathbf{r}) = \int_0^1 du\ q_{j} \hat{\mathbf{r}}_{j} \big[ 1
- u (\hat{\mathbf{r}}_j \cdot \nabla) + \frac{u^2}{2!} (\hat{\mathbf{r}}_j \cdot \nabla)^2 + \cdots  \big] \delta(\mathbf{r}) ,
\end{equation}
where $\nabla$ is the gradient operator. Inserting this expanded form of polarization operator in Eq.~\eqref{PZW_Ham}, we get
\begin{eqnarray} \label{Diracdelta}
    \hat{H}_{\mathrm{LM}} &&= - \sum_j \int_0^1 du\ q_{j} \int d^3 \mathbf{r} \ \hat{r}_{j}^\nu \big[ 1 - u (\hat{\mathbf{r}}_j \cdot \nabla) \nonumber \\ 
     && + \frac{u^2}{2!} (\hat{\mathbf{r}}_j \cdot \nabla)^2 \cdots  \big]  \delta(\mathbf{r}) E_\nu(\mathbf{r},t),
\end{eqnarray}
where $E_\nu(\mathbf{r},t)$ is the electric field in $\nu$-th direction. Here, we utilize the Einstein summation notation over Greek indices $\nu \ \epsilon \ \{x,y,z\}$ for the dot product, $a^{\nu}b_{\nu} = \sum_{\nu}a^{\nu}b_{\nu}$. We can use the identities of the Dirac delta distribution $ - \int d^3 \mathbf{r} \left(\hat{\mathbf r}_j \cdot \nabla\right) \delta(\mathbf r) f(\mathbf r) = (\hat{\mathbf r}_j \cdot \nabla) f(\mathbf 0)$ and so on in Eq.~\eqref{Diracdelta} and get
\begin{equation}
    \hat{H}_{\mathrm{LM}} = - \sum_j \int_0^1 du\ q_{j} \hat{r}_{j}^\nu \big[ 1
    + u (\hat{r}_j^{\kappa} \nabla_{\kappa}) + \cdots  \big] E_\nu(\mathbf{0},t).
\end{equation}
Upon integration over the variable $u$ it follows that \cite{Mukamel2009}
\begin{eqnarray} \label{fieldExpansion}
    \hat{H}_{\mathrm{LM}} =&& - \sum_j \bigg[q_j \hat{r}_j^{\nu} E_{\nu}(\mathbf{0},t) + \frac{1}{2!} q_j \hat{r}_j^{\nu} \hat{r}_j^{\kappa} \nabla_{\kappa} E_{\nu}(\mathbf{0},t) + \cdots \bigg]. \nonumber \\
     \equiv&& -\hat{\mu}^{\nu} E_{\nu}(\mathbf{0},t) -  \hat{Q}^{\nu \kappa} \nabla_{\kappa} E_{\nu}(\mathbf{0},t) + \dots ,
\end{eqnarray}
where $\hat{\mu}^{\nu}$ and $\hat{Q}^{\nu \kappa}$ represent the dipole and quadrupole moments, respectively.
The dipole approximation to the interaction Hamiltonian arises from truncating the series given in Eq.~\eqref{fieldExpansion} at the first term:
\begin{equation}    \label{HintDipole}
    \hat{H}^{\mathrm{dipole}}_{\mathrm{LM}} = - \hat{\mu}^{\nu} E_{\nu}(\mathbf{0},t).
\end{equation}
Higher-order multipolar terms can be used to correct the dipolar term, but the expansion convergence can be slow for highly structured fields, requiring an infinite number of terms in the expansion in the worst case scenario \cite{Horn2023}.

Furthermore, individual multipole moments beyond the lowest order depend on the choice of an expansion point in the Taylor series \cite{Lestrange2015}. As a result, finite-order truncations of the series can vary depending on the choice of an expansion point, even though the complete infinite series represents the same physical field. By replacing the expansion~\eqref{fieldExpansion} with Eq.~\eqref{PZW_Ham} we circumvent the issue of convergence and expansion point, and thus we can analyze the nonuniform light-matter interactions with the full spatial structure of the field.

In contrast to finite-order multipole expansions, the full interaction Hamiltonian for a charge-neutral system is intrinsically origin independent. In general, both the electronic and nuclear degrees of freedom couple to the electromagnetic field through their respective charge distributions. The total interaction term can be separated into an electronic contribution, arising from the coupling of the radiation field with the electronic density, and an ionic contribution, originating from the motion of the positively charged ions, composed of nuclei and tightly bound core electrons. However, when calculating light-matter interactions, it is common to consider \enquote{clamped} or \enquote{frozen} nuclei, where the external fields affect the electronic degrees of freedom of the system only. We have taken this approach here as it is easily interfaced with modern electronic structure calculations to parameterize the model. To  ensure that the system is electrically neutral and origin independent, we subtract the reference electronic density (see Sec.~\ref{chargebalance}).

\subsection{Wannier Basis}  \label{wannierbasis}
The computation of light-induced dynamics using Eq.~\eqref{PZW_Ham} has been limited by the challenge of computing the spatial integral at each time instance. To overcome this, we propose using a basis where the position operator is diagonal. Specifically, to efficiently capture the space-dependence of the light-matter interactions and to interface with the first-principle Hamiltonians, we adopt maximally-localized Wannier functions (MLWFs) as a basis set \cite{Marzari2012}.
These MLWFs are routinely extracted from first-principle computations of solids through a Boys localization procedure and used to develop generalized tight-binding Hamiltonians comparable in accuracy to the full plane-wave computation in a wide range of energies \cite{Marzari1997}. In this method, $N$ Wannier functions per unit-cell are used to describe a solid with $N$ bands. These basis functions $\{ |{n \mathbf{R}} \rangle \}$ with label $n = 1, \dots , N$ located in unit-cell with lattice vector $\mathbf{R}$ have the advantage of forming a highly localized and orthonormal basis set, i.e. $ \langle m \mathbf{R}'| n \mathbf{R} \rangle = \delta_{mn} \delta_{\mathbf{R} \mathbf{R}'} $. In this basis, the position operator matrix elements have a very simple form 
\begin{equation} \label{positionrep}
    \langle m \mathbf{R}' | \hat{\mathbf{r}} | n \mathbf{R} \rangle = \mathbf{R} \delta_{mn} \delta_{\mathbf{R}' \mathbf{R}} + \langle m (\mathbf{R}'-\mathbf{R})| \hat{\mathbf{r}} |n \mathbf{0} \rangle  .
\end{equation}
Here, the first term is the lattice vector $\mathbf{R}$ in the Wannier representation, which is diagonal, and the second term contains the inter- and intra-cell position matrix elements between the Wannier functions. The localization of Wannier functions shows that the overlap of these orbitals belonging to different unit-cells decreases exponentially with $|\mathbf{R}'-\mathbf{R}|$. Thus, we can often neglect the inter-cell matrix elements of $\hat{\mathbf{r}}$ and only consider the intra-cell contribution $D_{mn} = \langle m \mathbf{0}| \hat{\mathbf{r}} |n \mathbf{0} \rangle $ , simplifying Eq. \ref{positionrep}. We emphasize that this assumption is made for convenience and is not a necessary component of our proposed strategy, as shown explicitly in the Supplementary Material \cite{supmat} (Fig. S1). For now, we remark that, with this assumption, the resulting matrix is block-diagonal. Moreover, it is generally not possible to find a unitary transformation that diagonalizes Eq. \ref{positionrep} simultaneously along all cartesian coordinates. 
However, it  is possible to diagonalize \added{a cartesian component $\hat{\xi}$ of the position operator}, using a unitary transformation $U^{\xi}$,\added{ where $\xi\ \epsilon \ \{x,y,z\} $.} \deleted{the position operator \ref{positionrep} along a given direction $\xi$}

This transformation thus defines the \emph{modified} MLWFs $|\alpha_{\xi} \mathbf{R} \rangle = \sum_n U_{\alpha n}^{\xi} | n \mathbf{R} \rangle$. In this \emph{modified} MLWFs basis, the cartesian component \replaced{$\hat{\xi}$}{$\xi$} of the position operator is  diagonal and it is given by
\begin{equation}
    \langle \beta_{\xi} \mathbf{R}' |  \hat{\xi} | \alpha_{\xi} \mathbf{R} \rangle = ( R_{\xi} + \xi_{\alpha}) \delta_{\mathbf{R}' \mathbf{R}} \delta_{\alpha_{\xi} \beta_{\xi}} ,
\end{equation}
where, \added{$\xi_{\alpha}$ represents the eigenvalue} $\hat{\xi} | \alpha_{\xi} \mathbf{0} \rangle  = \xi_{\alpha} | \alpha_{\xi} \mathbf{0} \rangle $, \added{corresponding to the operator $\hat{{\xi}}$}. The advantage of this representation is that it greatly simplifies the computation of the multipole expansion to all orders, since powers of diagonal matrices are readily computed as $ \langle \beta_{\xi} \mathbf{R}' | { \hat{\xi}^n} |\alpha_{\xi} \mathbf{R} \rangle = (R_{\xi} + \xi_{\alpha} )^n \delta_{\mathbf{R}' \mathbf{R}} \delta_{\alpha_{\xi} \beta_{\xi}}$. We utilize this property to efficiently calculate the matrix elements of the multipolar PZW light-matter interaction Hamiltonian, as described in detail in the following section.

\subsection{Spatially-dependent light-matter interactions} \label{derivation}
The light-matter interaction term in Eq.~\eqref{PZW_Ham} is readily calculated using the \emph{modified} MLWFs. Any physical electric field $\mathbf{E}(\mathbf{r},t)$ can be decomposed numerically exactly into a finite sum of space-separable functions through an appropriate basis expansion as follows:
\begin{equation} \label{sum_field}
     \mathbf{E}(\mathbf{r},t) = \sum_i E_{ix}(x,t) E_{iy}(y,t) E_{iz}(z,t) \mathbf{e}_i,  
\end{equation}
where $\mathbf{e}_i$ denotes the polarization unit vector associated with the $i$-th term.
Thus, plugging Eq.~\eqref{sum_field} in Eq.~\eqref{PZW_Ham}, we can write $\hat{H}_{\mathrm{LM}}$ in second quantization as
\begin{widetext}
\begin{align} \label{HamSQ}
    \hat{H}_{\mathrm{LM}} &= \added{ \sum_{\substack{m,n \\ \mathbf{R}',\mathbf{R}}}  \langle m \mathbf{R}' | \bigg[ \int d^3 \mathbf{r} \int_0^1 du \ |e| \  \hat{\mathbf{r}} \ \delta[\mathbf{r} - u \hat{\mathbf{r}} ] \bigg] \cdot \bigg[ \sum_i E_{ix}(x,t) E_{iy}(y,t) E_{iz}(z,t) \mathbf{e}_i \bigg]  | n \mathbf{R} \rangle \hat{c}_{m \mathbf{R}'}^{\dagger}  \hat{c}_{n \mathbf{R}} } \nonumber \\
    &= |e| \sum_i \sum_{\substack{m,n \\ \mathbf{R}',\mathbf{R}}} \bigg( \langle m \mathbf{R}' | \int_0^1 du \int d^3 \mathbf{r} \ ( \hat{\mathbf{r}} \cdot \mathbf{e}_i ) \delta({x} - u \hat{{x}}) E_{ix}({x,t}) \delta({y} - u \hat{{y}}) E_{iy}({y,t})   \delta({z} - u \hat{{z}}) E_{iz}({z},t) | n \mathbf{R} \rangle \hat{c}_{m \mathbf{R}'}^{\dagger}  \hat{c}_{n \mathbf{R}} \bigg),
\end{align} 
where $-|e|$ is the electron charge and $\hat{c}_{m \mathbf{R}'}^{\dagger}$ $(\hat{c}_{n \mathbf{R}})$ creates (annihilates) a fermion in (from) state $| m \mathbf{R}' \rangle$ $(| n \mathbf{R} \rangle)$ . \deleted{For concreteness, we considered an electric field polarized along the $x$ direction, such that $\mathbf{e}_i = \mathbf{e}_x$, and $\hat{\mathbf{r}} \cdot \mathbf{e}_x = \hat{x}$. Furthermore,} 
\added{Now, for a general polarization direction, we have $\hat{\mathbf{r}} \cdot \mathbf{e}_i = \hat{r}_i$, where $\hat{r}_i$ is some combination of $\hat{x}$, $\hat{y}$ and $\hat{z}$. For notational simplicity we derive the expressions for an electric field polarized along the $x-$direction, such that $\hat{\mathbf{r}} \cdot \mathbf{e}_x = \hat{x}$. The corresponding expressions for $y-$ and $z-$polarized fields follows by replacing $\hat{x} \leftrightarrow \hat{y}$ and $\hat{x} \leftrightarrow \hat{z}$, respectively.}
\added{Furthermore,} we insert the resolution of identity in the \emph{modified} basis $ \hat{\mathbf{I}} = \sum_{\alpha_{\xi} \mathbf{R}} | \alpha_{\xi} \mathbf{R} \rangle \langle {\alpha_{\xi}} \mathbf{R} | $ in the above equation with the $\xi$ chosen as to simplify the action of the position operator along each of the $x, y$ and $z$ coordinate.
Hence, we can write the light-matter interaction term in Eq. \replaced{(\ref{HamSQ}) in the second quantization for an electric field polarized along the $x-$direction using the \emph{modified} basis}{ (\ref{PZW_Ham}) in second quantization} as  
\begin{align}
    \hat{H}_{\mathrm{LM}} &= |e| \sum_i \sum_{\substack{m,n \\ \mathbf{R}',\mathbf{R}}} \bigg(  \int_0^1 du  
            \int d {x} \sum_{\alpha_x, \mathbf{R}_1}  \langle m \mathbf{R}' | \hat{x} \   \delta({x} - u \hat{{x}}) | \alpha_x \mathbf{R}_1 \rangle E_{i\added{x}}({x,t}) \nonumber \\
            & \times \langle \alpha_x \mathbf{R}_1 | \int d y \sum_{\alpha_y, \mathbf{R}_2} \delta({y} - u \hat{{y}}) | \alpha_y \mathbf{R}_2 \rangle E_{i\added{y}}({y,t}) \nonumber \\
            & \times \langle \alpha_y \mathbf{R}_2 | \int d {z} \sum_{\alpha_z, \mathbf{R}_3} \delta({z} - u \hat{{z}}) | \alpha_z \mathbf{R}_3 \rangle E_{i\added{z}}({z},t) \langle \alpha_z \mathbf{R}_3 | n \mathbf{R} \rangle  \bigg) \hat{c}_{m \mathbf{R}'}^{\dagger}  \hat{c}_{n \mathbf{R}} 
            \nonumber \\
            &= |e| \sum_i \sum_{\substack{m,n \\ \mathbf{R}',\mathbf{R}}} \bigg(  \int_0^1 du  
            \sum_{\alpha_x, \mathbf{R}_1}  \langle m \mathbf{R}'|  \alpha_x \mathbf{R}_1 \rangle ( R_{1x} + {x}_{\alpha} ) \int d{x}\ \delta \big({x} - u (R_{1x} + {x}_{\alpha} ) \big)  E_{i\added{x}}({x,t}) \nonumber \\
            & \times \sum_{\alpha_y, \mathbf{R}_2} \langle \alpha_x \mathbf{R}_1 | \alpha_y \mathbf{R}_2 \rangle \int d{y}\  \delta \big({y} - u (R_{2y} + {y}_{\alpha} ) \big) E_{i\added{y}}({y,t}) \nonumber \\
            & \times \sum_{\alpha_z, \mathbf{R}_3}  \langle \alpha_y \mathbf{R}_2 | \alpha_z \mathbf{R}_3 \rangle \int d{z}\ \delta \big({z} - u (R_{3z} + {z}_{\alpha} ) \big) E_{i\added{z}}({z},t) \langle \alpha_z \mathbf{R}_3 |  n \mathbf{R} \rangle  \bigg) \hat{c}_{m \mathbf{R}'}^{\dagger}  \hat{c}_{n \mathbf{R}}.
\end{align}
To further simplify, we can use the property of Dirac delta function, $ \int ds\ \delta( s - a) f(s) = f(a) $. This gives
\begin{align} \label{Hint_Pol}
        \hat{H}_{\mathrm{LM}} &= |e| \sum_i \sum_{\substack{m,n \\ \mathbf{R}',\mathbf{R}}}\ \sum_{\substack{\alpha_x,\alpha_y,\alpha_z \\ \mathbf{R}_1,\mathbf{R}_2,\mathbf{R}_3}}  \bigg(  \int_0^1 du \langle m \mathbf{R}'|  \alpha_x \mathbf{R}_1 \rangle ( R_{1x} + {x}_{\alpha} )  E_{i\added{x}}\big[u (R_{1x} + {x}_{\alpha}),t  \big] \langle \alpha_x \mathbf{R}_1 | \alpha_y \mathbf{R}_2 \rangle \nonumber \\
        & \times \ E_{i\added{y}}\big[u (R_{2y} + {y}_{\alpha} ),t \big] \langle \alpha_y \mathbf{R}_2 | \alpha_z \mathbf{R}_3 \rangle E_{i\added{z}}\big[u (R_{3z} + {z}_{\alpha} ),t \big]   \langle \alpha_z \mathbf{R}_3 | n \mathbf{R} \rangle  \bigg) \hat{c}_{m \mathbf{R}'}^{\dagger}  \hat{c}_{n \mathbf{R}} ,
\end{align}
where $ \langle \alpha_z \mathbf{R}_3 | n \mathbf{R} \rangle $ corresponds to the overlap integral between the Wannier basis and the diagonal basis in the $z$ direction, and so on.
Equation~\eqref{Hint_Pol} defines the general light-matter interaction in MLWFs basis, and the form of electric field includes all the multipolar terms in a compact form which can be easily applied in computations. \added{Moreover, the corresponding expressions for $y-$ and $z-$polarized fields are obtained by substituting the $R_{1x} + {x}_{\alpha}$ 
from the polarization to 
$R_{2y} + {y}_{\alpha} $ and $R_{3z} + {z}_{\alpha} $, respectively, with the electric field components unchanged.}
Similarly, the light-matter interaction term within the EDA [Eq.~\eqref{HintDipole}] for an electric field polarized in the $x$ direction can be written using MLWFs as 
\begin{equation}
    \label{Hint_dipole}
    \hat{H}^{\mathrm{dipole}}_{\mathrm{LM}} = |e| \sum_{\substack{m,n \\ \mathbf{R}',\mathbf{R}}}  \langle m \mathbf{R}' | \hat{x} \ E(0,t) | n \mathbf{R} \rangle \hat{c}_{m \mathbf{R}'}^{\dagger}  \hat{c}_{n \mathbf{R}}  = |e| \sum_{m,n, \mathbf{R}} ( R_x \delta_{mn} + \langle m \mathbf{0}| \hat{x} |n \mathbf{0} \rangle ) E(0,t) \hat{c}_{m \mathbf{R}}^{\dagger}  \hat{c}_{n \mathbf{R}} .
\end{equation} 
\end{widetext}

We note that if magnetic-field effects are significant then the leading-order magnetic multipolar expansion terms can be added independently to Eq~\eqref{Hint_Pol}. In the PZW formulation, the magnetization operator $\mathbf{\hat{M}}(\mathbf{r}) \sim \mathbf{\hat{r}} \times \mathbf{\hat{p}}$ \cite{Mukamel2009}, such that the magnetic interaction depends simultaneously on the position and momentum operators. Because $\mathbf{\hat{r}}$ and $\mathbf{\hat{p}}$ do not commute, they cannot be simultaneously diagonalized. As a result, one cannot construct a localized  basis for the magnetic interaction term, unlike the treatment of polarization operator in Eq.~\ref{PZW_Ham}. While magnetic multipolar expansion terms can be expressed straightforwardly in the MLWFs representation, finite-order truncation of the multipole expansion may reintroduce origin dependence \cite{Lestrange2015}.

\subsection{Equation of motion} \label{EqOfMotion}

For effective single-particle Hamiltonians like those obtained from DFT, the electronic properties are completely determined by the single-particle electronic reduced density matrix
\begin{equation}
    \rho_{n \mathbf{R},m \mathbf{R}'}(t) = \langle \Psi(t)| \hat{c}_{n  \mathbf{R}}^{\dagger}  \hat{c}_{m \mathbf{R}'} |\Psi(t)\rangle \equiv \langle \hat{c}_{n \mathbf{R}}^{\dagger}  \hat{c}_{m \mathbf{R}'} \rangle ,
\end{equation}
where $|\Psi(t)\rangle$ is the many-body wavefunction. The dynamics of $\rho_{n \mathbf{R},m \mathbf{R}'}(t)$  is governed by the Liouville-von Neumann equation
\begin{equation} \label{rhodt}
    i \hbar \frac{d}{dt} \rho_{n \mathbf{R},m \mathbf{R}'}(t) = \langle [ \hat{c}_{n \mathbf{R}}^{\dagger}  \hat{c}_{m \mathbf{R}'} , \hat{H}(t ) ] \rangle ,
\end{equation}
with initial conditions $ \rho_{n \mathbf{R},m \mathbf{R}'}(0) = \langle \Psi(0)| \hat{c}_{n \mathbf{R}}^{\dagger}  \hat{c}_{m \mathbf{R}'} |\Psi(0) \rangle$. In quantum dynamics, it is useful to employ an orbital decomposition for $\rho_{n \mathbf{R},m \mathbf{R}'}(t)$. Let $|\epsilon \rangle$ be eigenorbitals for the material i.e., $\hat{H}_\mathrm{M} |\epsilon \rangle = \epsilon |\epsilon \rangle$. Using the basis transformation $\hat{c}^{\dagger}_{n \mathbf{R}} = \sum_{\epsilon} \langle \epsilon | n \mathbf{R} \rangle \hat{c}^{\dagger}_{\epsilon}$, the initial single-particle electronic reduced density matrix at initial time can be expressed as
\begin{equation} \label{initialrho}
    \rho_{n \mathbf{R},m \mathbf{R}'}(0) = \sum_{\epsilon,\epsilon'} \langle \epsilon | n \mathbf{R} \rangle \langle m \mathbf{R}'| \epsilon' \rangle \langle \Psi(0)| \hat{c}^{\dagger}_{\epsilon} \hat{c}_{\epsilon'} |\Psi (0) \rangle,
\end{equation}
where $ \hat{c}^{\dagger}_{\epsilon}$ $(\hat{c}_{\epsilon})$ creates (annihilates) a fermion in (from) the orbital $|\epsilon \rangle$ with energy $\epsilon$. For an initial single-determinant wavefunction $\langle \Psi(0)| \hat{c}^{\dagger}_{\epsilon} \hat{c}_{\epsilon'} |\Psi (0) \rangle = \delta_{\epsilon,\epsilon'} f(\epsilon)$, where $f(\epsilon) = 1$ for filled level and 0 otherwise. The quantity characterizes the initial electronic distribution among the single-particle states.
Upon time evolution, the matrix element $\rho_{n \mathbf{R},m \mathbf{R}'}(t)$ maintains the form
\begin{equation} \label{rho_nm(t)}
    \rho_{n \mathbf{R},m \mathbf{R}'}(t) = \sum_{\epsilon} \langle \epsilon (t)| n \mathbf{R} \rangle \langle m \mathbf{R}'| \epsilon(t) \rangle f(\epsilon).
\end{equation}
The utility of using the orbital decomposition is that if the time-dependent orbitals $|\epsilon(t)\rangle$ satisfy the single-particle Schr\"odinger equation
\begin{equation} \label{eom}
    i \hbar \frac{d}{dt} |\epsilon(t) \rangle = \hat{H}(t) |\epsilon(t) \rangle,
\end{equation}
with initial condition $|\epsilon(0) \rangle$, the single-particle electronic reduced density matrix automatically satisfies the correct equation of motion [Eq.~\eqref{rhodt}]. 

\section{Computational approach} \label{model}

\subsection{Model system}   \label{model system}

To explore light-matter interactions beyond the dipole approximation, we employ a tight-binding model Hamiltonian of a 1-dimensional solid with parameters obtained using the electronic structure calculations of \emph{trans}-polyacetylene (\emph{t}PA). Specifically, to obtain MLWFs we perform density functional theory computations using Quantum ESPRESSO (QE) \cite{Giannozzi2009} with the Perdew-Zunger parametrization \cite{Perdew1981} for the exchange-correlation functional, employing an ultra-soft pseudopotential\cite{DalCorso2014}, and a plane wave cutoff of 100 Ry. We use the geometry of \emph{t}PA structure reported in Ref. \cite{Ferretti2012}, with bond length alternations of 1.34/1.54 \r{A}, and a unit-cell length of $a=2.496$ \r{A}. 
We then perform a Wannier interpolation of the DFT electronic structure using the Wannier90 package \cite{Mostofi2014}. We parametrize the highest valence and lowest conduction band of bulk \emph{t}PA and construct a generalized tight-binding model with 2 MLWFs per unit cell with 6 nearest-neighbor couplings. The generalized tight-binding Hamiltonian can be written as 
\begin{equation} \label{Hmatter}
 \hat{H}_{\mathrm{M}} = \sum_{\substack{m,n \\ \mathbf{R}',\mathbf{R}}} h^{mn}_{\mathbf{R}'\mathbf{R}} \ \hat{c}_{m \mathbf{R}'}^{\dagger}  \hat{c}_{n \mathbf{R}} ,
\end{equation}
where $h^{mn}_{\mathbf{R}'\mathbf{R}}$ are the matrix elements obtained from Wannier90. The Hamiltonian parameters lead to a energy gap between the highest occupied molecular orbital (HOMO) and lowest unoccupied molecular orbital (LUMO) of $1.68$ eV for a finite but extended system. 
We diagonalize the matter Hamiltonian~\eqref{Hmatter} to obtain the eigenorbitals. The system is then initialized in an equilibrium state with orbitals below the Fermi energy being filled and orbitals above the gap being empty. The single-particle density matrix is propagated in time according to Eq.~\eqref{eom} using the Adams-Moulton predictor-corrector method with adaptive time step in the SUNDIALS package \cite{Hindmarsh2005}.\\

\subsection{Incorporating charge balance} \label{chargebalance}
We now remark on the nature of the light-matter interaction in Eq.~\eqref{PZW_Ham}. The nanostructure is overall charge-neutral with negative electronic charges balanced by the positive nuclear charges. However, the MLWFs basis consists of electronic orbitals only, obtained from the electronic structure problem under the assumption of fixed nuclei. As a result, the nuclear contribution to charge neutrality is not explicitly represented in the MLWFs description.
To enforce charge neutrality at the operator level and ensure origin-independent dynamics, we subtract from Eq.~\eqref{Hint_Pol} the reference electronic density associated with the initial single-particle reduced density matrix [Eq.~\eqref{initialrho}]. This subtraction, implemented in Eq.~\eqref{HDiag_exact}, removes the static background charge from the field coupling so that the interaction depends only on charge fluctuations about the neutral reference state. The resulting dynamics are invariant under shifts of the coordinate origin, as verified explicitly in the Supplemental Material \cite{supmat} (Fig. S2). 

The charge-neutral multipolar light-matter interaction Hamiltonian from Eq.~\eqref{Hint_Pol} can then be written directly using \emph{modified} MLWFs basis $\{ |\alpha_{x} \mathbf{R} \rangle \}$ along the polarization direction as
\begin{widetext}
\begin{align} \label{HDiag_exact}
         \hat{H}_{\mathrm{LM}} &= |e| \int_0^1 du \sum_i \sum_{\substack{\alpha_x,\beta_x \\ \mathbf{R}_1,\mathbf{R}'_1}} \sum_{\substack{\alpha_y,\alpha_z \\ \mathbf{R}_2,\mathbf{R}_3}} \bigg( ( R'_{1x} + {x}_{\beta} )  E_{ix}\big[u (R'_{1x} + {x}_{\beta} ), t \big] \langle \beta_x \mathbf{R}'_1 | \alpha_y \mathbf{R}_2 \rangle E_{iy}\big[u (R_{2y} + {y}_{\alpha} ), t \big]  \nonumber \\
         & \times \langle \alpha_y \mathbf{R}_2 | \alpha_z \mathbf{R}_3 \rangle E_{iz}\big[u (R_{3z} + {z}_{\alpha} ),t \big] \langle \alpha_z \mathbf{R}_3 | \alpha_x \mathbf{R}_1 \rangle \bigg) \bigg[ \hat{c}^{\dagger}_{\beta_x \mathbf{R}'_1} \hat{c}_{\alpha_x \mathbf{R}_1}  - \rho_{\beta_x \mathbf{R}'_1,\alpha_x \mathbf{R}_1}(0) \delta_{\beta_x,\alpha_x} \delta_{ \mathbf{R}'_1, \mathbf{R}_1} \bigg] .
\end{align}
Throughout, we employ Eq.~\eqref{HDiag_exact} in our simulations. In the dipole approximation, Eq.~\eqref{HDiag_exact} reduces to
\begin{equation} \label{HDiag_dipole}
H^{\mathrm{dipole}}_{\mathrm{LM}} = |e|
\sum_{\alpha_x \mathbf{R} } ( R_x + {x}_{\alpha} ) E(0,t) \bigg[ \hat{c}^{\dagger}_{\alpha_x \mathbf{R}}  \hat{c}_{\alpha_x \mathbf{R}} -  \rho_{\alpha_x \mathbf{R},\alpha_x \mathbf{R}}(0) \bigg].    
\end{equation} 

\end{widetext}

\subsection{Numerical details}  \label{computations}

In the localized Wannier basis, the interaction Hamiltonian assumes a block-diagonal structure for each unit cell labeled by lattice vector $\mathbf{R}$ when the inter-cell contributions are neglected as described in Sec.\ref{wannierbasis}. For the minimal case of 2-MLWFs per unit cell, Eq.~\eqref{HDiag_exact} can be evaluated directly. For more than 2-MLWFs per unit cell, the integral over $u$ in Eq.~\eqref{HDiag_exact} can be evaluated numerically using Gauss–Legendre quadrature \cite{Golub1969,Hale2013}\replaced{, which requires}{ .After} mapping the integration domain $u \ \epsilon\ [0,1]$ \cite{Abramowitz2013}\replaced{. The}{, the} interaction Hamiltonian Eq.~\eqref{HDiag_exact} reduces to 
\begin{widetext}
\begin{eqnarray} \label{Hint_GQ}
         \hat{H}^{\mathrm{GQ}}_{\mathrm{LM}} \approx && |e| \sum_{k=1}^{N_q} \sum_i \sum_{\substack{\alpha_x,\beta_x \\ \mathbf{R}_1,\mathbf{R}'_1}} \sum_{\substack{\alpha_y,\alpha_z \\ \mathbf{R}_2,\mathbf{R}_3}} w_k \bigg( ( R'_{1x} + {x}_{\beta} )  E_{ix}\big[u_k (R'_{1x} + {x}_{\beta} ), t \big] \langle \beta_x \mathbf{R}'_1 | \alpha_y \mathbf{R}_2 \rangle E_{iy}\big[u_k (R_{2y} + {y}_{\alpha} ), t \big]  \nonumber \\
         && \times \langle \alpha_y \mathbf{R}_2 | \alpha_z \mathbf{R}_3 \rangle E_{iz}\big[u_k (R_{3z} + {z}_{\alpha} ),t \big] \langle \alpha_z \mathbf{R}_3 | \alpha_x \mathbf{R}_1 \rangle \bigg) \bigg[ \hat{c}^{\dagger}_{\beta_x \mathbf{R}'_1} \hat{c}_{\alpha_x \mathbf{R}_1}  - \rho_{\beta_x \mathbf{R}'_1,\alpha_x \mathbf{R}_1}(0) \delta_{\beta_x,\alpha_x} \delta_{ \mathbf{R}'_1, \mathbf{R}_1} \bigg] ,
\end{eqnarray}
\end{widetext}
where $u_k$ and $w_k$ are quadrature nodes and weights, and $N_q$ is the number of quadrature points which control numerical accuracy. The nodes $u_k$ are given by the roots of the Legendre polynomial $P_{N_q}(u)$, and the corresponding weights $w_k$	are computed using the standard expressions provided in Ref.~\cite{Abramowitz2013}.
The quadrature is introduced solely as a numerical convenience and does not modify the formal structure of the theory.

To probe the limits of dipole approximation, we want to perform the computations in a parameter regime where the total length of the chain is comparable to the the incident wavelength $\lambda$ or the illuminated area. The challenge is that this requires computations that are beyond what is presently achievable. For instance, for a resonant excitation of $\hbar \omega = 1.68$ eV (corresponding to the wavelength of $\lambda = 738$ nm), we need about 3200 unit cells resulting in a Hamiltonian of dimension $6400 \times 6400$ with a unit-cell length of $a=2.496$ \r{A}.
Therefore, to access the parameter regime of interest we artificially scale the position matrix elements $ ( R_{\xi} + \xi_{\alpha}) $ by a factor $\gamma>1$. In turn, the field amplitude is divided by the same factor $E_0 \rightarrow E_0/\gamma$, ensuring that the dipole component of the light-matter interaction term of the Hamiltonian [Eq.~\ref{HDiag_dipole}] remains unchanged.

We characterize the dynamics caused by the light-matter interaction by analyzing the average energy $\langle \hat{H}(t) \rangle = \mathrm{Tr} \big[ \hat{H}(t) \ \hat{\rho}(t) \big]$, and the photoinduced polarization along the laser's polarization $\langle \hat{P}(t) \rangle = \mathrm{Tr} \big[  \hat{{x}} \  \hat{\rho}(t) ) \big]$. We also monitor orbital occupation, given by the single-particle density matrix populations in the energy basis $ \rho_{\epsilon \epsilon} $. To compare the dynamics generated by the multipolar PZW vs. the EDA Hamiltonian, we calculate the difference in energy absorbed by the system from the laser field $\Delta \langle \hat{H} \rangle = \big(\langle \hat{H}(t_{\mathrm{f}}) \rangle - \langle \hat{H}(t_0) \rangle \big) $, where $t_0$ and $t_{\mathrm{f}}$ are times before and after the laser pulse, respectively. The final time is chosen well after the pulse when the light-matter term in the Hamiltonian is zero. 

To quantify the performance of the dipole and low-order multipolar corrections, we also calculate the fidelity between the polarization generated by the approximate and the full multipolar light-matter interaction terms. This fidelity is computed as
\begin{equation} \label{Fidelity}
    F(f(t),g(t)) =  \frac{ \big| \int_0^T f^*(t) g(t) dt \big|}{ ||f|| \ ||g|| },
\end{equation}
where $T$ is the total duration and $ ||f|| = \sqrt{ \int_0^T |f(t)|^2 dt  }$ is the normalization constant  \cite{Nielsen2012}. It is a quantity between 0 and 1, which quantifies the overlap between two time-dependent functions over a time interval $T$, providing a measure of how closely the polarization obtained under the dipole approximation aligns with the multipolar PZW solution.

\begin{figure*}[htb]
\centering
\includegraphics[width=1.0\textwidth, keepaspectratio]{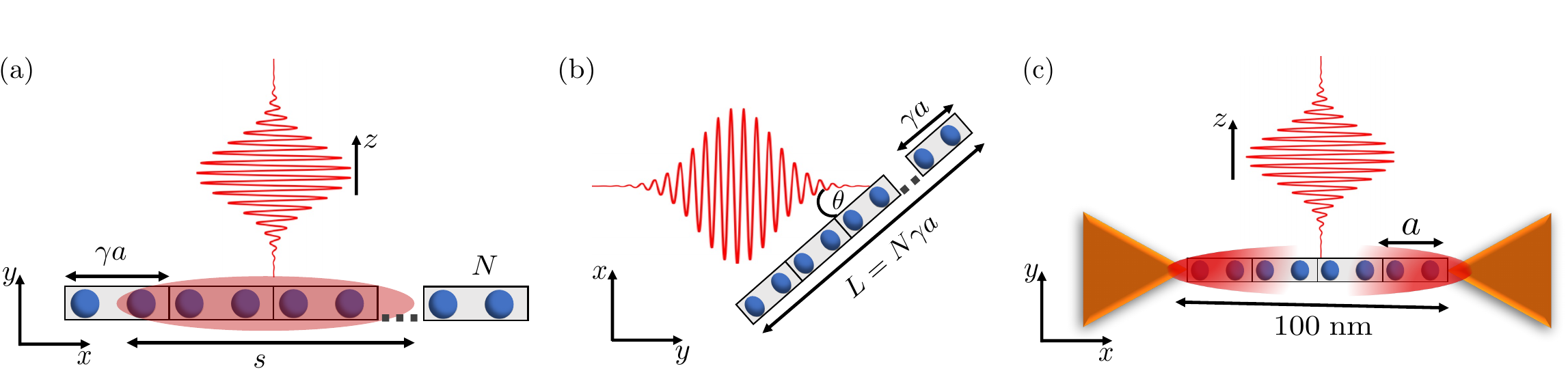}
\caption{Schematic illustrations of the three light-matter interaction scenarios considered in this work. The 1-D chain is (a) illuminated by a Gaussian (in space and time) beam that propagates perpendicular to the chain with a spot size $s$ smaller than the chain length $L=N\gamma a$; (b) tilted at an angle $\theta$ relative to the $y$ direction of propagation of a Gaussian (in time) pulse, polarized along the $x$ axis; (c) placed in a $100$ nm gap of a bow-tie metal-dielectric-metal junction and illuminated by a Gaussian (in time) pulse that propagates perpendicular to the chain.}
\label{fig:diagrams}
\end{figure*}

\subsection{Light-matter interaction scenarios considered} \label{scenarios}

\subsubsection{Non-uniform illumination} \label{subNonuniform}

To examine the limitation of the dipole approximation in capturing a nonuniform electric field profile we consider a chain that is partially illuminated as shown in Fig. \ref{fig:diagrams} (a).
The chain is illuminated by the Gaussian beam that propagates along the $z$ direction and is polarized along the chain direction (in $x$). 
The electric field generated by this laser beam is given by
\begin{align}    \label{GaussianBeam}
    \mathbf{E}(x,y,z,t) & = \mathrm{Re} \bigg\{ E_0  \exp{\bigg[-\frac{((t-t_0)-z/c)^2}{\sigma^2}\bigg]}  \nonumber \\
    & \exp{\bigg[-\frac{x^2 + y^2}{w(z) ^2}\bigg]} \exp{\big[i(kz-\omega (t-t_0))\big]} \mathbf{e}_x \bigg\}. 
\end{align}
Here $\omega$ is the central frequency of the laser, $E_0$ its amplitude, $\sigma$ the temporal width of the laser envelope, $t_0$ the center of the temporal envelope, $w(z)$ the beam radius, $c$ the speed of light, and $\mathbf{e}_x$ the polarization unit vector \cite{Siegman1986}. The quantity $k = 2 \pi/\lambda$ is the magnitude of the wavevector, where $\lambda$ is the wavelength and $\omega = k c$. 

The chain is contained in the $xy$-plane with $z=0$ and no dipole matrix elements along the $z$ direction. Thus, Eq.~\eqref{GaussianBeam} simplifies to 
\begin{align} \label{gaussian}
    \mathbf{E}(x,y,z=0,t) &= E_0 \exp\bigg[-\frac{(t-t_0)^2}{\sigma^2}\bigg] \exp \bigg[- \frac{x^2 + y^2}{s^2}\bigg] \nonumber \\
    & \cos \big( \omega (t-t_0) \big) \mathbf{e}_x ,
\end{align}
where $s=w(0)$ is the spot size of the beam at $z=0$. Conveniently, this field can be written as a single term in the sum [Eq.~\eqref{sum_field}] with time-independent $E_x(x) = \exp[-x^2/s^2]$, $E_y(y) = \exp[-y^2/s^2]$, and no $z$-dependence $E(t) = E_0  \exp{\big[-(t-t_0)^2/\sigma^2\big]} \cos{ \big( \omega (t-t_0) \big)}  $. 

The parameters of the Gaussian beam used in the simulations are as follows. The central frequency is set to match the band gap of the material, $\hbar \omega = 1.68$ eV, corresponding to the wavelength of $\lambda = 738$ nm. The pulse has a temporal width $\sigma = 20$ fs, centered around $t_0 = 80$ fs. In calculations presented in the Sec. \ref{nonuniform}, the spot size is kept at a constant value of $s = 800$ nm, which is greater than the diffraction limit \cite{Siegman1986}. The scaling parameter of the unit-cell is set to $\gamma = 10$ so that the length of a chain composed of $N$ sites is $L= N \gamma a$.

\subsubsection{Pushing the long-wavelength limit} \label{subLongwavelength}

To explore the limits of the long-wavelength condition, we consider a generalized scenario shown in Fig \ref{fig:diagrams} (b) with the 1-D chain tilted with respect to the direction of propagation of the electric field at an angle of $\theta$. In this case the Gaussian pulse propagates along the $y$ direction with the polarization along the $x$ axis and the chain is contained in the $xy$ plane ($z=0$). The electric field is therefore given by
\begin{equation} \label{gaussianY}
    \mathbf{E}(x,y,z=0,t) = E_0 \exp\bigg[-\frac{(t-t_0)^2}{\sigma^2}\bigg] \cos \big( k y - \omega (t-t_0) \big) \mathbf{e}_x ,
\end{equation}
and, as in Sec.~\ref{subNonuniform}, can be represented by a single term of the sum~\eqref{sum_field}. In addition to changing the chain and the field propagation direction relative to Sec.~\ref{subNonuniform}, the spatial envelope of the electric field is removed, corresponding to a spot size that is much larger than the size of the system. All other parameters are the same as in Sec.~\ref{subNonuniform}.

\subsubsection{Beyond-dipole effects with a realistic electric field}\label{realistic-setup}

Lastly, as shown in Fig. \ref{fig:diagrams} (c), we consider a problem with a realistic electric field in a complex electromagnetic environment created by illuminating a bow-tie metallic junction with a Gaussian laser pulse. The bow-tie antenna geometry is well known for producing strongly enhanced and spatially non-uniform electromagnetic fields localized in the nanoscale gap \cite{Nien2013,Khalil2021}.
To obtain the electric field, we solved Maxwell's equations numerically using the Finite-Difference Time-Domain (FDTD) method, with spatial discretizations on the Yee grid \cite{Taflove2005} utilizing the MEEP (MIT Electromagnetic Equation Propagation) software package \cite{Oskooi2010}. 
The details of the electromagnetic simulation are as follows. We consider a plane wave of a wavelength $\lambda = 738 $ nm with a Gaussian envelope of a temporal width $\sim 20.8$ fs, interacting with a bow-tie metal-dielectric-metal nanojunction. The laser pulse propagates along the $z$ direction and is linearly polarized along the junction in the $x$-direction. 
\added{We set the grid resolution as $70, 100,\text{ and } 120$ points per $\mu\mathrm{m}$ and observe electric field profile within $1 \%$ of difference for the latter two resolutions.
Thus, for our simulations,} the grid resolution is set to 100 points per $\mu\mathrm{m}$, and time steps $\Delta t = S \Delta x/ c$ are used, where $\Delta x$ is the spacing between the grid points, $c$ is the speed of light and $S=0.3$ is the Courant factor used for numerical stability in FDTD simulations. 

The bow-tie antenna consists of two identical perfectly conducting metallic triangular prisms arranged tip-to-tip with a gap of $w = 100$~nm in a symmetric configuration. The structure has a principal axis passing through the prism tips and a vertical mirror plane bisecting the center of the junction. Each prism is defined in the \textit{xy}-plane by three vertices: one tip located at $(\pm w/2, 0)$ and two base vertices extending across the full width of the simulation domain along the \textit{y}-direction, forming an isosceles triangle. The prisms are extruded along the \textit{z}-direction to a height of 3~$\mu$m, corresponding to the physical thickness of the structure in the simulation. The computational cell is cubic with $3~\mu\mathrm{m}$ side length and surrounded by 1.5~$\mu\mathrm{m}$ of perfectly matched layers (PML) \cite{Berenger1994} to absorb outgoing radiation and prevent unphysical reflections caused by simulation cell boundaries. The dielectric region between the prisms is modeled as a uniform medium with relative permittivity $\varepsilon_r = 1$, corresponding to vacuum. In this scenario, we obtain a highly non-uniform field in the gap due to enhancement at the sharp metallic tips, which highlights the importance of beyond-dipole effects at the metal-dielectric interfaces. 

Unlike in Secs.~\ref{subNonuniform}-\ref{subLongwavelength}, the lattice constant is not scaled ($\gamma=1$). The chain consists of $N=400$ unit-cells, which gives the chain length $L=Na \sim 100$ nm. The spatial profile of the electric field $E(x,t)$ obtained from the simulation grid was interpolated onto the positions of the unit cell sites using cubic spline interpolation. We contrast the dynamics generated by the full PZW Hamiltonian, Eq.~\eqref{HDiag_exact}, and its dipole approximation [Eq.~\eqref{HDiag_dipole}] as well as the low-order multipolar corrections (up to octupole). The expansion is performed at the origin, so that the dipole term and its low-order corrections are calculated at the center of the junction.

\section{Results} \label{results}

\subsection{Nonuniform illumination} \label{nonuniform}

\begin{figure}[htb]
\centering
\includegraphics[width=0.45\textwidth,keepaspectratio]{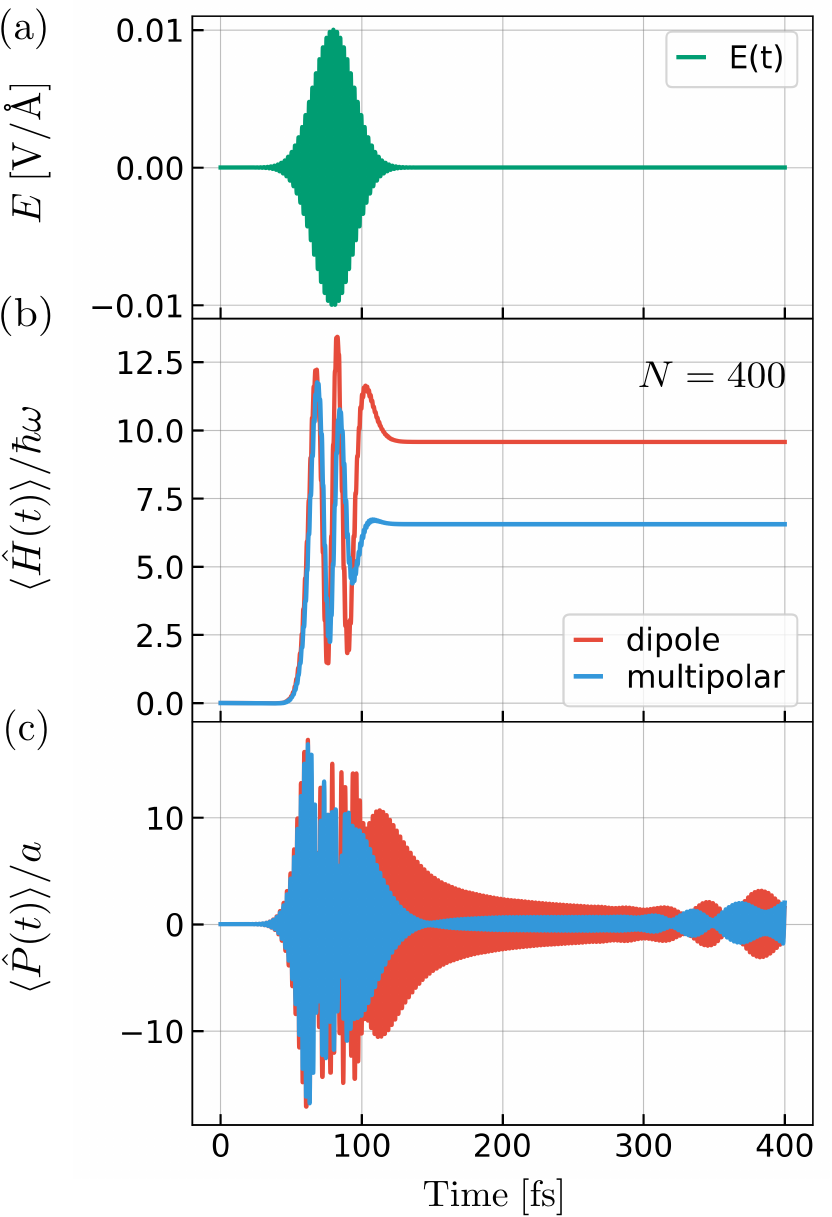}
\caption{Resonant excitation of the 1-$\mu$m-long 1-D chain illuminated with a Gaussian beam [Eq.~\eqref{gaussian}] with a spot size $s$ of 0.8 $\mu$m as in the geometry of Fig.~\ref{fig:diagrams}(a). (a) Temporal profile of the laser field $E(t)$. (b) The average energy and (c) polarization of the  chain with $N=400$ unit cells comparing the true dynamics generated by the multipolar PZW (blue) Hamiltonian to the dipole approximated Hamiltonian (red).}
\label{fig:AvgE}
\end{figure}

\begin{figure}[htb]
\centering
\includegraphics[width=0.45\textwidth,keepaspectratio]{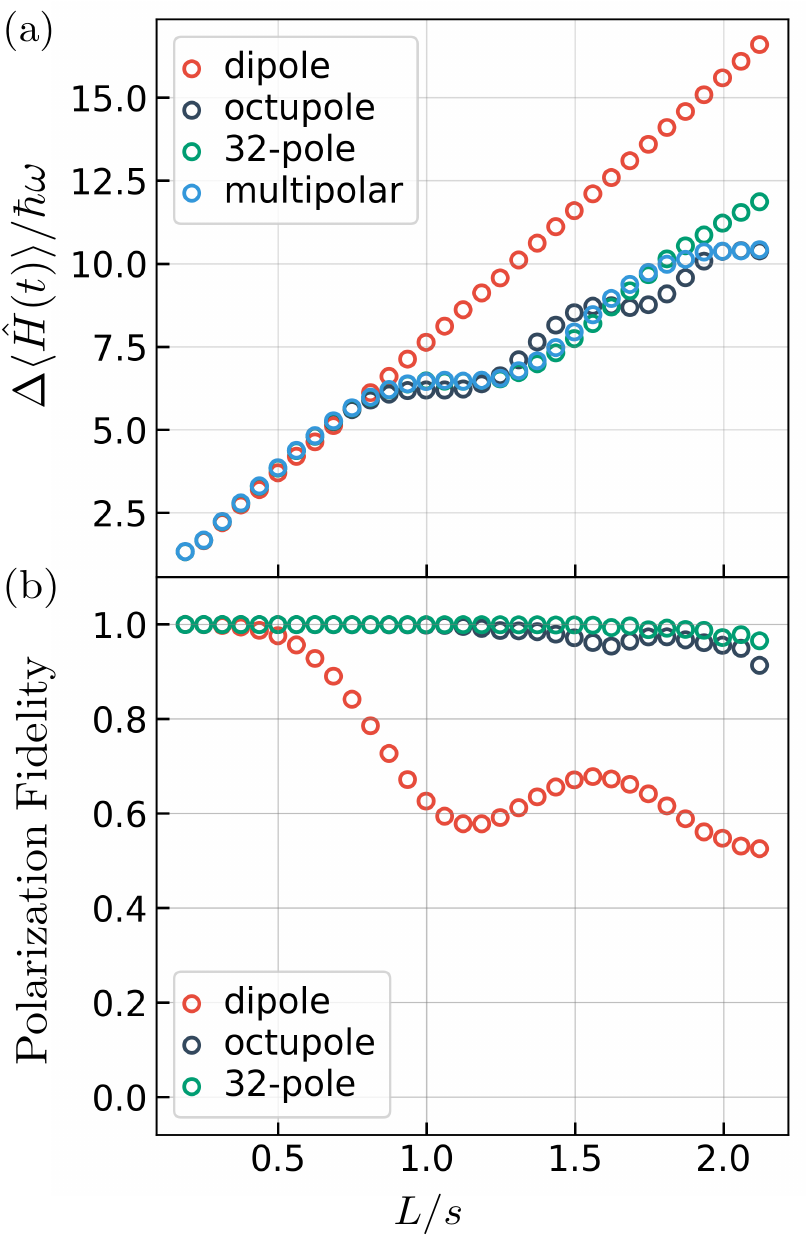}
\caption{(a) Net energy absorbed after photoexcitation $\Delta \langle \hat{H} \rangle$ and (b) the polarization fidelity [Eq.~\eqref{Fidelity}] for varying chain length $L$ with respect to a fixed spot size $s$ as in the geometry of Fig. \ref{fig:diagrams}(a).  Multipolar result is shown in blue, dipole in red, octupole in black and 32-pole in green. Simulation parameters are the same as in Figure~\ref{fig:AvgE}, except $N$ is allowed to vary between 80 and 700 unit cells. }
\label{fig:Ncomparison}
\end{figure}

\begin{figure}[htb]
\centering
\includegraphics[width=0.45\textwidth,keepaspectratio]{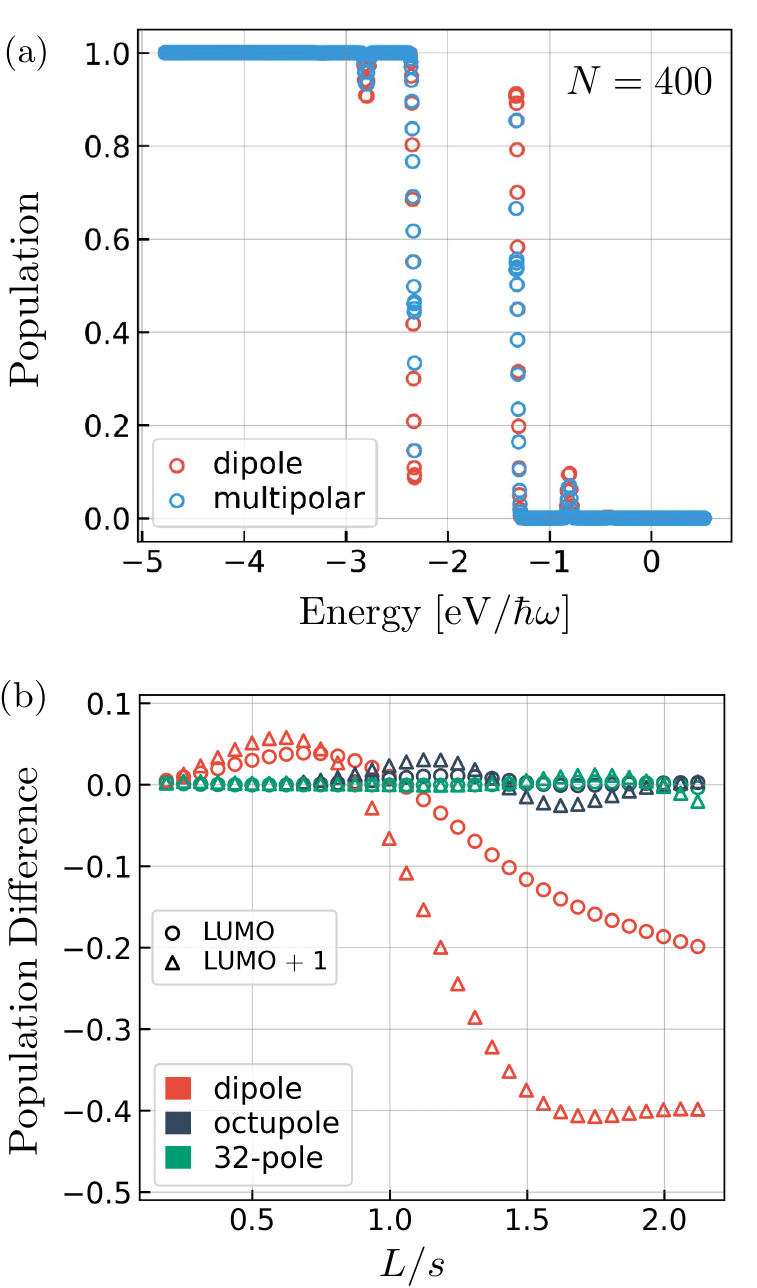}
\caption{Comparison of orbital populations after resonant excitation between the PZW multipolar (blue) and its finite-order approximations as in the geometry of Fig. \ref{fig:diagrams}(a). (a) Orbital populations for a chain with $N=400$. (b) Population difference between the full multipolar Hamiltonian and the finite-order corrections for states around the HOMO-LUMO gap as a function of chain length. The color coding and simulation parameters are the same as in Fig.~\ref{fig:Ncomparison}.}
\label{fig:population}
\end{figure}

We first examine the limitation of the dipole approximation to capture a nonuniform electric field profile by considering a chain that is partially illuminated as shown in Fig. \ref{fig:diagrams}(a). The results of such a simulation are shown in Fig.~\ref{fig:AvgE}. 
The total length of the chain $L=1000$ nm is larger than the spot size $s=800$ nm, corresponding to $L/s=1.25$. We monitor the photoinduced dynamics due to resonant photoexcitation through the total energy absorbed  $\Delta \langle \hat{H} \rangle$. The temporal profile of the laser is shown in Fig.~\ref{fig:AvgE}(a). We demonstrate in Fig.~\ref{fig:AvgE}(b)  that the dipole approximation overestimates the amount of energy absorbed by the system. This is to be expected, as the Gaussian beam is assumed to have a uniform spatial intensity, whereas in reality the edges of the chain are weakly illuminated relative to the middle of the chain. The photoinduced polarization along the chain axis shown in Fig.~\ref{fig:AvgE}(c) displays a smaller amplitude relative to the dipole approximation when the spatial structure of light is taken into account. With Fig.~\ref{fig:AvgE} we thus demonstrate that the dipole approximation is not accurate when the material is illuminated partially.

To demonstrate that the proposed scheme inherits the desirable convergence properties with Hilbert space dimension of the length  gauge in dipole approximation, we explicitly verified the convergence with respect to the number of Wannier orbitals. Specifically, we compared the laser-induced dynamics yielded by models with two and six MLWFs per unit cell considering the nonuniform illumination of 1-D chain as in Sec.\ref{subNonuniform} (see Fig. S3 in the Supplementary Material \cite{supmat}), showing identical results. This contrasts with computations in the velocity gauge that typically requires a large number of bands to yield converged results~\cite{Schueler2021,Tiwari2025}.

To further characterize this limitation of the dipole approximation, we investigate the photoinduced dynamics of chains with varying system size $L$ while keeping the beam spot size $s$ constant. We also investigate whether low-order multipolar terms (up to 32-polar) are sufficient to correct the dynamics (see Supplementary Material \cite{supmat} for explicit forms). As shown in Sec IV of Supplementary Material \cite{supmat}, for this laser field, the quadrupolar and other even order terms vanishes due to symmetry around the beam center. Fig.~\ref{fig:Ncomparison}(a) shows the net energy absorbed after photoexcitation, and panel (b) shows the polarization fidelity [Eq.~\eqref{Fidelity}] for the total duration $T=400$ fs of the dipole and higher-order approximations with respect to the full multipolar PZW Hamiltonian.

For $L < \frac{1}{2}s$, approximate results predict identical absorption and demonstrate polarization fidelity greater than 92\%. The limitations of the dipole approximation become evident when the system size $L$ becomes comparable with the spot size $s$. For $L > s$, the dipole approximation starts to significantly overestimate the net absorption of energy (Fig.~\ref{fig:Ncomparison}a) and provides an inaccurate description of the dynamics (Fig.~\ref{fig:Ncomparison}b). Adding the higher-order correction terms recovers the correct energy absorption and dynamics for larger values of $L/s$.  

Figure~\ref{fig:population} shows the orbital populations at $200$ fs, that is, well after the resonant photoexcitation with the Gaussian beam. Fig.~\ref{fig:population}(a) focuses on the same system as in Fig.~\ref{fig:AvgE}. The dipole approximation qualitatively captures the excitation profile with population peaks spaced by $\hbar\omega$ and $2\hbar\omega$ around the band gap, but overestimates the amount of population transfer. In Fig.~\ref{fig:population}(b) we plot the population difference between the multipolar PZW and the finite-order approximations as a function of $L/s$ for two orbitals with energy right above the HOMO-LUMO gap. Differences in the HOMO and HOMO-1 populations are not shown as they are a mirror image of the LUMO and LUMO+1, respectively. Echoing Fig.~\ref{fig:Ncomparison}, the populations differences are small up to $L\approx s/2$ with higher-order corrections generally improving the accuracy of the dipole approximation.

\subsection{Validity of Electric-Dipole Approximation}

\begin{figure*}[htb] 
\centering
\includegraphics[width=1.0\textwidth,keepaspectratio]{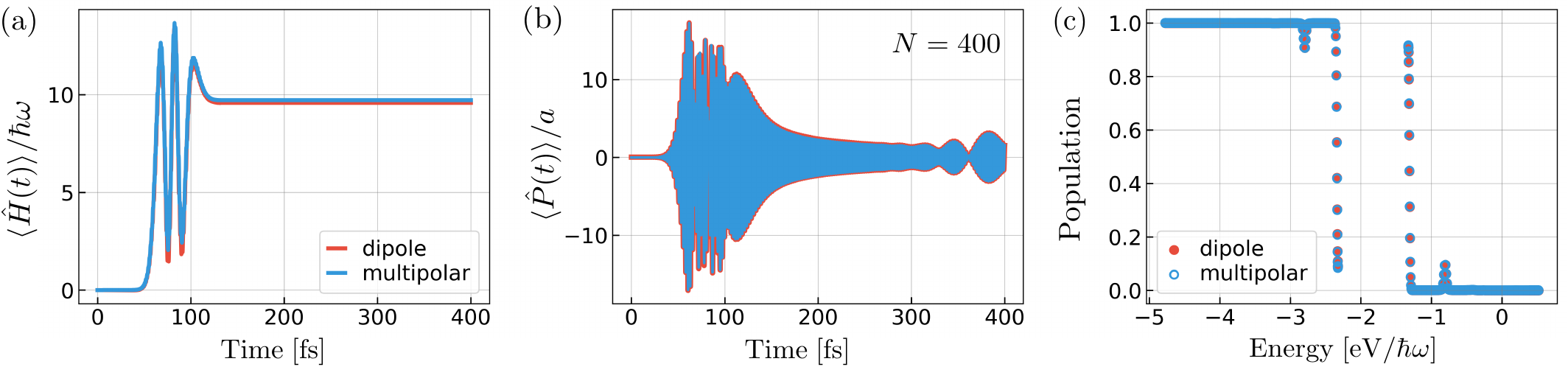}
\caption{The average energy (a), polarization (b) and population (c) of the  chain with $N=400$ unit cells comparing the dynamics generated by the multipolar PZW (blue) Hamiltonian to the dynamics obtained with dipole approximation (red). Simulation parameters are the same as in Figure~\ref{fig:AvgE}, except $s=10$ $\mu$m.} 
\label{fig:AvgElambda}
\end{figure*}
The EDA assumes a spatially uniform field. Thus, it is expected to hold when (i) the sample is illuminated uniformly and simultaneously (ii) the wavelength of light $\lambda$ is significantly larger than the sample size (long-wavelength limit) \cite{CohenTannoudji1997}. In the previous section, we explored the breakdown of EDA due to violations of (i). In this section, we ensure that condition (i) holds and investigate EDA's validity with respect to (ii).

However, violating (ii) does not necessarily invalidate the dipole treatment. For example, for periodic solids, the EDA holds when the wavelength $\lambda$ is long with respect to the unit-cell length $a$ (i.e. $\lambda \gg a$) and not with respect to the material's size \cite{Chernyak1995,Mukamel2009,Yu2010}. Here we show that for nanoscale materials, the EDA holds when  $\lambda$ is significantly larger than the size of the material \emph{along the direction of propagation of light}. Crucially, the other two dimensions can be arbitrarily large without compromising the EDA.

Figure~\ref{fig:AvgElambda} exemplifies this scenario numerically. The spot size $s=10$ $\mu$m is much larger than the system size $L=1$ $\mu$m, so the chain is illuminated uniformly. The wavelength $\lambda=738$ nm is smaller than the chain length but much larger than the thickness of the chain ($L_{z}\sim a$) along the direction of light propagation. In this case, we observe that the dynamics generated by the multipolar PZW Hamiltonian and the dipole approximation are identical.

Why does the wavelength need to be long with respect to the material dimension \textit{only} along the propagation direction of light?
This weaker validity condition for the  dipole approximation follows immediately from the fact that light is a transverse wave. Thus, the electric field variation in space due to the wave nature of light is captured entirely by the phase factor $e^{i\mathbf{k}\cdot\mathbf{r}}$. It is the magnitude of this phase factor that needs to be small for the dipole approximation to hold, that is, the requirement is $|\mathbf{k}\cdot\mathbf{r}|\ll1$. However, the magnitude of the $\mathbf{k}\cdot\mathbf{r}$ term is determined only by the projection of $\mathbf{r}$ onto $\mathbf{k}$ and its variation depends only on the coordinate along $\mathbf{k}$. The size of the sample along dimensions perpendicular to the direction of propagation does not have a bearing on the validity of the dipole approximation. This distinction is especially powerful for 1-D or 2-D material illuminated by perpendicular light beams. In this case, as our analysis shows, the EDA is accurate regardless of the material's size. We used this in Sec.~\ref{nonuniform} to isolate the effects of non-uniform illumination.

\subsection{Pushing the long-wavelength limit}
\begin{figure}[htb]
\centering
\includegraphics[width=0.45\textwidth,keepaspectratio]{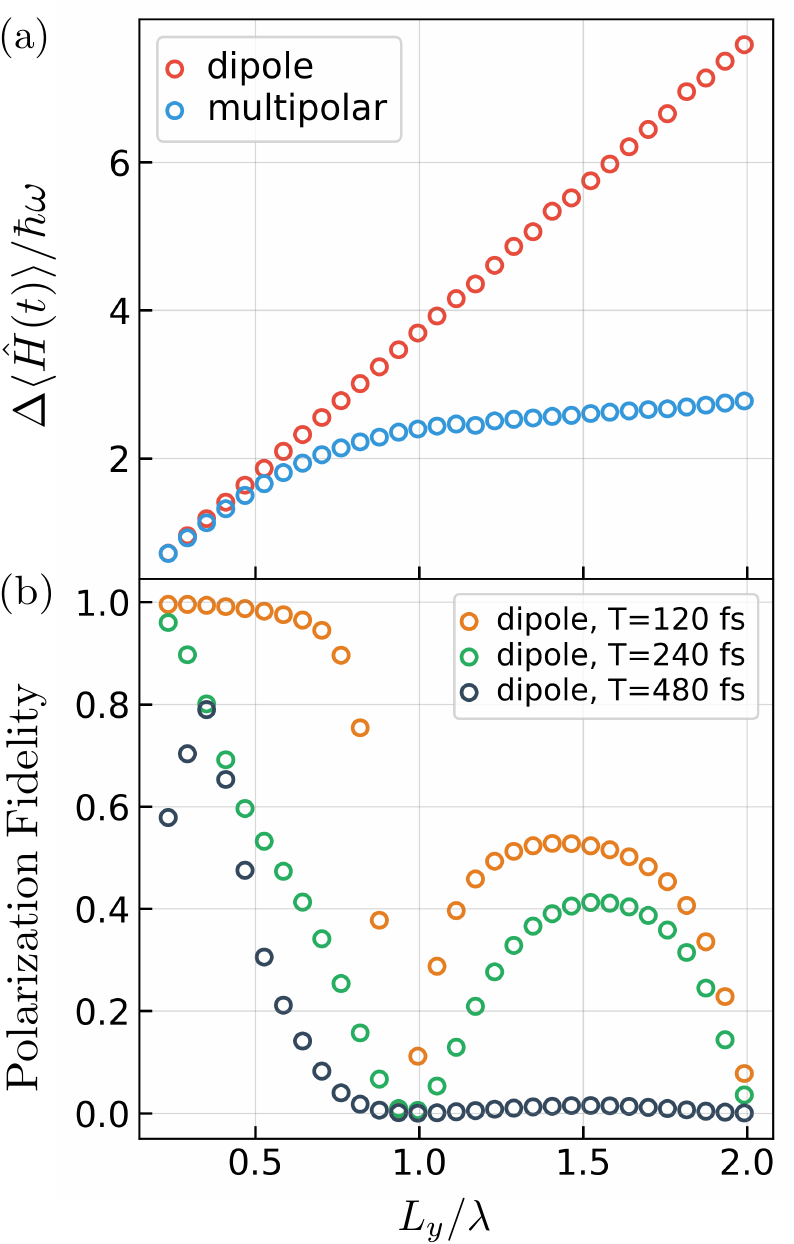}
\caption{The average energy (a) and polarization fidelity (b) for the chain tilted at $\theta=\pi/6$ (Fig.~\ref{fig:diagrams}b) for varying chain lengths with $N$ varying between  80 and 680 unit cells. The x axis is based on the system size along the polarization direction ($L_y$). The pulse spot size $s$ is taken to be infinite to ensure uniform spatial illumination. Other simulation parameter are identical to Fig.~\ref{fig:Ncomparison}. }
\label{fig:systemvswavelength}
\end{figure}

To make effects beyond the long-wavelength limit salient for a low-dimensional system, one thus needs to tilt it by an angle $\theta$ with respect to the direction of propagation of the electric field. 
Figure \ref{fig:systemvswavelength} presents the results for tilting our 1-D chain by angle $\theta = \pi/6$, as depicted in Fig.~\ref{fig:diagrams}(b). In Figs.~\ref{fig:systemvswavelength} (a) and (b), we show the total energy absorbed during the resonant excitation and the polarization fidelity, respectively, as a function of the linear dimension of the material along the direction of propagation of light ($L_y$) relative to the wavelength ($\lambda$). The dipole approximation becomes inaccurate due to violating the long-wavelength condition when the electric field varies over the system.

More specifically, the computations reveal that the EDA captures the early-time light-induced dynamics reasonably well until the system size along the direction of propagation $L_y$ is $\sim30 \%$ of the wavelength. To make this evident,
figure~\ref{fig:systemvswavelength}(b) shows the polarization calculated over three different time durations. It specifically demonstrates that early-time polarization dynamics (in orange) is captured exceptionally well until when $L_y/\lambda$ of $\sim30\%$. It is also worth noting that the polarization fidelity does not decrease monotonically with $L_y/\lambda$. The decay of fidelity appears to be modulated by an oscillation with a period of $~1$. Upon comparing the induced polarization of the dipole and the multipolar case, we conclude that the polarization in the multipolar case goes out of phase with respect to the dipole as it approaches  $L_y \sim \lambda $. For $\lambda < L_y < 1.5\lambda$, the overlap between dipole and multipolar polarization increases due to closer matching of the amplitude of the oscillations, but then decreases again as amplitude for multipolar decreases.
We remark that in experimental settings it might be difficult to maintain coherent dynamics long after the laser pulse. Instead, the inevitable decoherence will quickly dampen the polarization oscillations, making the fidelity calculated over $T=120$ fs the most experimentally relevant. As such, we expect the dipole approximation to provide an excellent description of laser-induced dynamics up to $L_y/\lambda$ of $\sim30\%$, assuming that the decoherence effects that dampen the polarization oscillations are accounted for.

\begin{figure}[htb]
\centering
\includegraphics[width=0.48\textwidth,keepaspectratio]{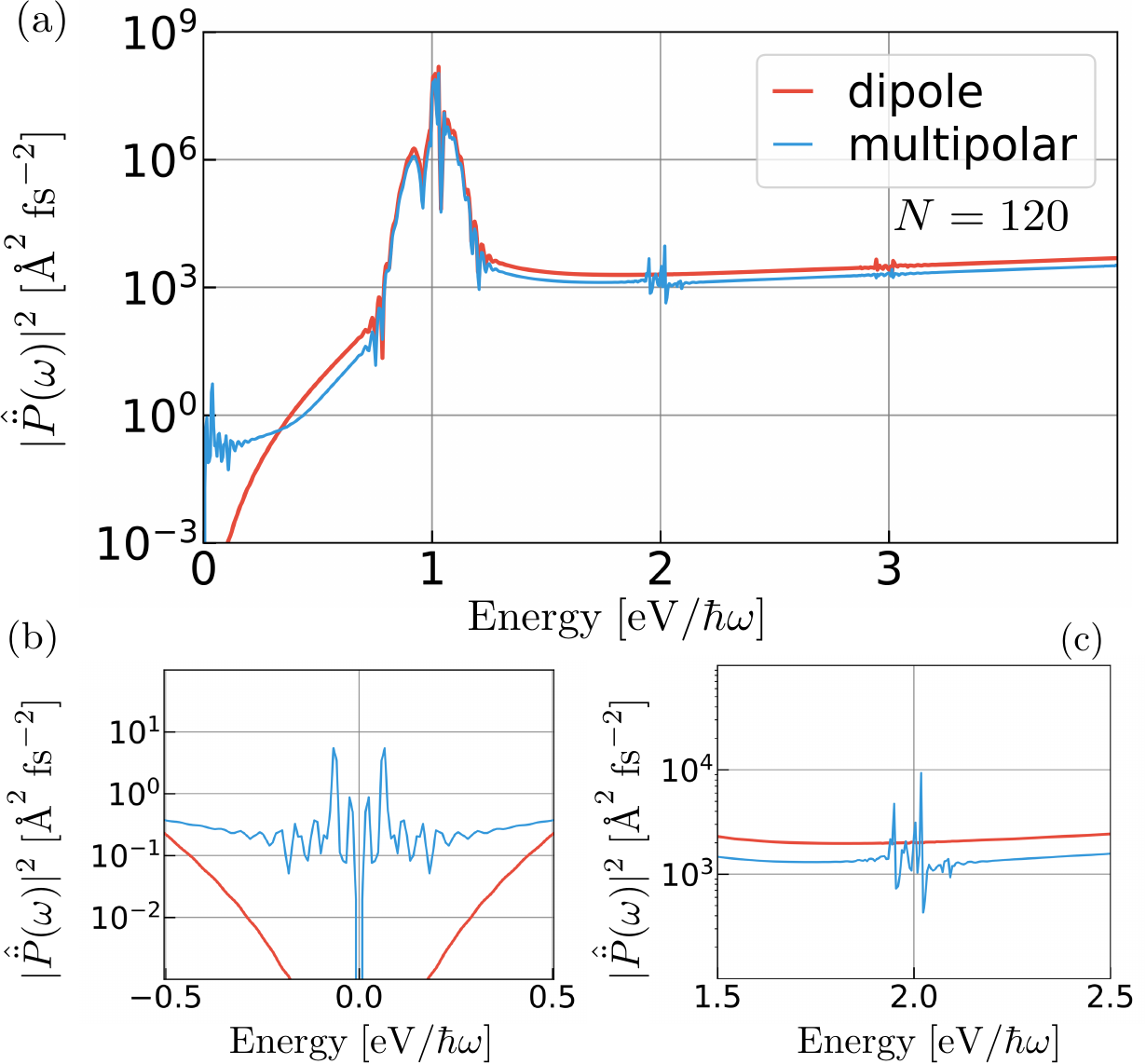}
\caption{Power spectra of polarization acceleration for $N = 120$ unit-cells, corresponding to $L_y=260$ nm with other simulation parameters as in Fig.~\ref{fig:systemvswavelength}. Panel (a) shows the full spectrum, whereas (b) and (c) zoom in to the zero frequency and second-harmonic peaks, respectively.}
\label{fig:harmonics}
\end{figure}

\begin{figure}[htb]
\centering
\includegraphics[width=0.48\textwidth,keepaspectratio]{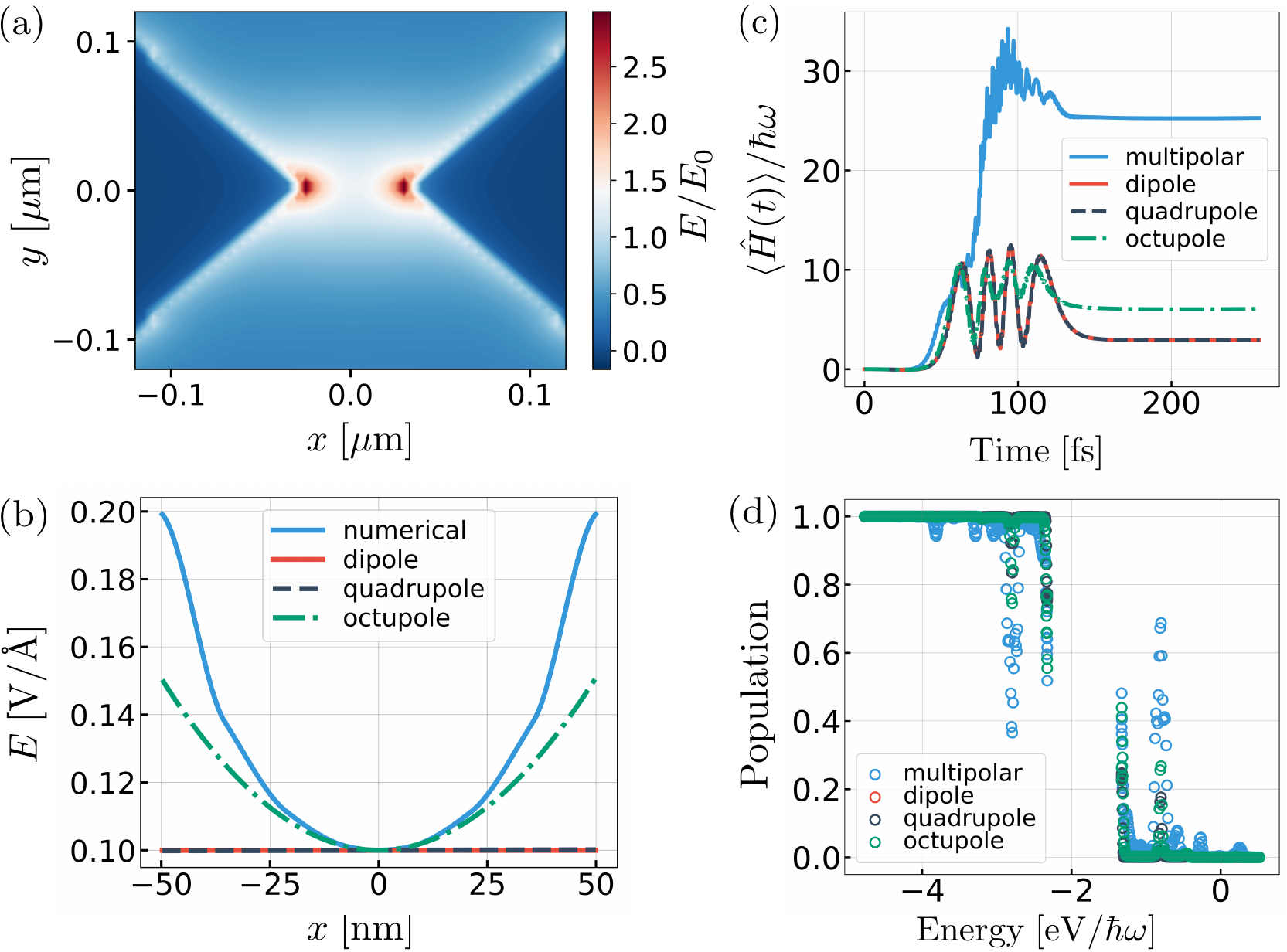}
\caption{Dynamics of the 1-D chain placed in a gap of a bow-tie metal-dielectric-metal junction and illuminated by a perpendicular Gaussian (in time) pulse. (a) $z=0$ cut showing the magnitude of the electric field at time $t_0 = 80$ fs and demonstrating its enhancement due to the sharp metallic edges. (b) Electric field amplitude along the 1-D chain ($y=2.5, z=0$, $t=t_0$) shown in blue and its approximations with upto a given finite-order multipolar term. (c) Average energy with respect to time and (d) excited-state population after the pulse. The 100-nm long chain contains $N=400$ unit cells.} 
\label{fig:electrodynamicsim}
\end{figure}

\begin{figure*}[htb] 
\centering
\includegraphics[width=1\textwidth, keepaspectratio]{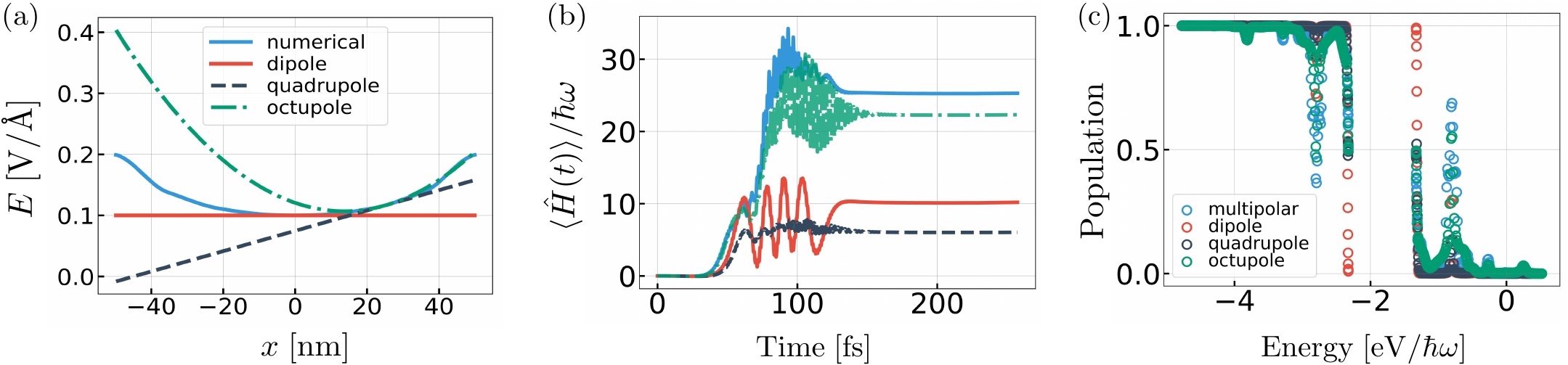}
\caption{(a) Electric field amplitude along the 1-D chain at an off-center expansion point ($x= 25$ nm). (b) Average energy with respect to time and (c) excited-state population after the pulse for $N=400$ unit cells.} 
\label{fig:Randompoint}
\end{figure*}
We next present the power spectra obtained from the Fourier transform of polarization acceleration $|\mathcal{F} \{ \langle \hat{\ddot{P}}(t) \rangle \}|^2 = | -\omega^2 P(\omega) |^2 $ (Fig.~\ref{fig:harmonics}) to demonstrate the ability to predict harmonic generation beyond the EDA with our methodology. We observe that the first and third harmonic peaks are comparable for both methods, as shown in Fig.~\ref{fig:harmonics}(a). By contrast, Fig.~\ref{fig:harmonics} (c) clearly demonstrates that the multipolar dynamics captures a strong second harmonic peak, which is absent in the EDA. This is because even harmonics are forbidden in inversion symmetric systems (like ours) under the EDA. However, in reality, due to spatial variations of the electric field, inversion symmetry is effectively broken, an effect that is naturally captured within our multipolar framework. 
In addition, as shown in Fig.~\ref{fig:harmonics}(b), the dipole approximation strongly suppresses the low-frequency response near $\omega \approx 0$. By contrast, multipolar dynamics exhibits pronounced low-frequency peaks, reflecting the emergence of difference-frequency generation between nearby optical frequency components. These low-frequency components arise from coherent intra-band electron motion and correspond to the emission of terahertz radiation following resonant optical excitation.

These findings echo previous studies demonstrating that spatially structured fields from optical near fields \cite{Iwasa2009,Noda2017} or including non-dipole higher-order corrections for strong field emission \cite{Mishra2012,Jensen2025} can recover the even-order harmonic generation that is missing in the dipole approximation. We emphasize that our method captures these beyond-dipole phenomena for weak intensity far-field light and does so computationally efficiently, i.e. at the cost comparable to dipole calculations.

We also note an open question regarding the validity of the electric–dipole approximation in extended systems. For the finite nano- and microscale structures studied here, the relevant length scale is the total system size $L$, and the long-wavelength limit requires $L \ll \lambda$. In contrast, in periodic materials described by Bloch’s theorem, the unit-cell size $a$ becomes the relevant length scale of the material \cite{Chernyak1995,Mukamel2009,Yu2010}, allowing the dipole approximation to remain valid in bulk solids. In the context of our simulations, this implies that upon increasing $L_y$ in Fig.~\ref{fig:systemvswavelength}, the dipole approximation should ultimately converge towards the exact result, indicating a crossover from a finite-size extended system to a periodic bulk regime. However, reaching this asymptotic regime remains beyond our current computational capabilities.

\subsection{Beyond-dipole effects with realistic electric field}

The calculations presented in previous sections were performed with the idealistic models of electric fields, given via analytical forms of their spatial and temporal components [Eqs.~\eqref{gaussian} and \eqref{gaussianY}]. Here we consider a realistic electric field from an electrodynamic simulation of the pulse illuminating the bow-tie junction, as described in Sec.~\ref{realistic-setup} and Fig.~\ref{fig:diagrams}(c). We emphasize that this set-up is common for the characterization of laser-driven petahertz electronics \cite{Dombi2013,Rybka2016,Khalil2021,Heide2024}. Typically, these junction-based setups manifest a spatial structure of the electric field that differs significantly from light propagation in vacuum. This is because the sharp metallic tips significantly distort the incident light. Here we explore the result of this added spatial structure of the electric field on the performance of the ubiquitous dipole approximation.

Figure~\ref{fig:electrodynamicsim} (a) clearly shows the enhancement of the electric field $E/E_0$ close to the sharp tips of the bow-tie junction, where $E_0 = 0.1$ V/\r{A} is the incident field strength from the source and $E$ is the field amplitude at $t_0 = 80$ fs. The simulation parameters are detailed in Sec.~\ref{scenarios}. Figure~\ref{fig:electrodynamicsim}(b) compares the electric field at time~$t_0$ as given by the dipole approximation (red) and its low-order multipolar corrections with the full spatial profile (blue). Because the field is almost symmetric near the center of the junction, it is commonly used as the expansion point in the EDA. The low-order corrections fail to reproduce the spatial non-uniformity present in the full field. The first derivative is zero at this expansion point. Thus, the energy and population dynamics with the quadrupolar correction [Fig.~\ref{fig:electrodynamicsim}~(c-d)] remain essentially identical to those obtained under the dipole approximation alone. The octupolar correction captures the features of the electric field near the expansion point but fails to fully reproduce the accurate dynamics. Fig.~\ref{fig:electrodynamicsim}(c) and (d) demonstrate higher final energy and more intense population peaks in the multipolar description relative to the finite-order approximations, indicating that they significantly underestimate the strength of the excitation.

Before concluding, we wish to highlight a key limitation of finite-order expansions: the dipole and its corrections are sensitive to the chosen expansion point in the Taylor series.
Figure~\ref{fig:Randompoint} (a) shows the field profile of the dipole approximation (red) and its finite-order multipole corrections at an off-center point, as well as the full spatial profile of the realistic electric field (blue). Fig.~\ref{fig:Randompoint} (b) and (c) shows the resulting energy and population dynamics at the expansion point which are quite different from Fig.~\ref{fig:electrodynamicsim}. 
 As this example demonstrates, selecting a different expansion point can drastically change the approximate electric field profile, and consequently lead to markedly inconsistent dynamical predictions.
In contrast, the full multipolar Hamiltonian avoids the numerical issues of the finite-order expansion and provides correct light-induced dynamics without significantly increasing the computational cost relative to the dipole approximation. 

\subsection{Computational cost of beyond-dipole approximation}

\added{The computational cost of the PZW Hamiltonian remains of the same order as that of the dipole approximation. We consider two cases. First, when the electric field is separable into spatial and temporal components [Eq.~\ref{gaussian}], the temporal dependence in Eqs.~\ref{HDiag_exact} and~\ref{Hint_GQ} can be factored out. The spatial part of the interaction Hamiltonian is computed once and subsequently multiplied by the time-dependent field, resulting in a computational cost nearly identical to that of the dipole approximation (see Supplementary Material \cite{supmat}, Fig.~S4(a)).}
\added{Second, when the spatial and temporal components are not separable [Eq.~\ref{gaussianY}], the interaction Hamiltonian must be evaluated at every time step. Since Eq.~\ref{HDiag_exact} is block diagonal (neglecting inter-cell contributions) and therefore sparse, we exploit this structure using optimized sparse matrix operations. As shown in Supplementary Material \cite{supmat}, Fig.~S4(b), this introduces only a modest computational overhead. In both cases, the numerical wavefunction propagation dominates the runtime, so the overall computational cost remains comparable to that of the dipole approximation.}

\section{Conclusions} \label{conclusion}

We have developed a general theoretical framework for light-matter interactions beyond the dipole approximation, applicable to nano- and micro-scale materials interacting with arbitrary electric fields without truncation at finite multipolar order. The full multipolar PZW interaction Hamiltonian is expressed in the \emph{modified} MLWFs basis that diagonalizes the position operator to facilitate calculations. This framework provides a computationally efficient methodology for simulating quantum dynamics in extended systems that interacts with an arbitrary spatially structured electric field.

We used this newly developed framework to characterize the limits of the ubiquitous dipole approximation. By isolating the individual effects of the nonuniform illumination vs. the breakdown of the long-wavelength limit, we quantified their respective bounds of applicability. The long-wavelength limit breaks down when the relevant linear dimension of the material reaches $\sim30\%$ of the light's wavelength $\lambda$. Crucial for studies of low-dimensional materials, we found that the dipole approximation remains accurate for uniformly illuminated 1-D and 2-D materials, provided the light propagates perpendicular to the low-dimensional material even when the system size is much larger than the wavelength of the light along the confined dimensions.
Independently from the long-wavelength limit, the dipole approximation also fails if the illumination is non-uniform over the system. For a specific case of the Gaussian beam, this failure becomes noticeable when the material becomes comparable to the spot size. Adding the finite-order correction terms can recover the correct dynamics for smoothly varying Gaussian fields but fails completely for the highly non-homogeneous metal-molecule junction case and are sensitive to the expansion point. We also observe beyond-dipole effects dictated by symmetry, such as the generation of even-order harmonics and zero frequency peaks when the system with inversion symmetry interacts with spatially structured light.

To summarize, our newly developed formalism for light-matter interactions beyond the dipole approximation enables incorporation of the full spatial structure of the electric field to calculate light-induced dynamics in extended systems of nano- and microscopic size, including fields with complex spatial structures, such as those found in hetero-junctions, at the same computational cost of a standard dipole calculation. This efficiency paves the way for accurate first-principles simulations of spatially-structured light-matter dynamics in nanoscale devices, quantum materials, and interfaces.

\begin{acknowledgments}
    This study is based on work supported by the National Science Foundation under Grant No. CHE-2416048.
\end{acknowledgments}

\section*{Data Availability}

The numerical data supporting the main-text figures,
together with the plotting
scripts, are available in the Figshare repository
\cite{figshare_data}. Additional numerical data,
including those supporting the Supplemental Material,
and the simulation source code are available from
the authors upon reasonable request.

\bibliography{bibliography}

@article{Condon1937,
  title = {Theories of Optical Rotatory Power},
  author = {Condon, E. U.},
  journal = {Rev. Mod. Phys.},
  volume = {9},
  issue = {4},
  pages = {432--457},
  numpages = {0},
  year = {1937},
  month = {Oct},
  publisher = {American Physical Society},
  doi = {10.1103/RevModPhys.9.432},
  url = {https://link.aps.org/doi/10.1103/RevModPhys.9.432}
}

@Article{Heide2024,
  author    = {Heide, Christian and Keathley, Phillip D. and Kling, Matthias F.},
  journal   = {Nature Reviews Physics},
  title     = {Petahertz electronics},
  year      = {2024},
  issn      = {2522-5820},
  month     = oct,
  number    = {11},
  pages     = {648--662},
  volume    = {6},
  doi       = {10.1038/s42254-024-00764-7},
  publisher = {Springer Science and Business Media LLC},
}

@Article{Dombi2013,
  author    = {Dombi, Péter and Hörl, Anton and Rácz, Péter and Márton, István and Trügler, Andreas and Krenn, Joachim R. and Hohenester, Ulrich},
  journal   = {Nano Letters},
  title     = {Ultrafast Strong-Field Photoemission from Plasmonic Nanoparticles},
  year      = {2013},
  issn      = {1530-6992},
  month     = jan,
  number    = {2},
  pages     = {674--678},
  volume    = {13},
  doi       = {10.1021/nl304365e},
  publisher = {American Chemical Society (ACS)},
}

@Article{Ferretti2012,
  author    = {Ferretti, Andrea and Mallia, Giuseppe and Martin-Samos, Layla and Bussi, Giovanni and Ruini, Alice and Montanari, Barbara and Harrison, Nicholas M.},
  journal   = {Phys. Rev. B},
  title     = {Ab initiocomplex band structure of conjugated polymers: Effects of hydrid density functional theory and GW schemes},
  year      = {2012},
  issn      = {1550-235X},
  month     = jun,
  number    = {23},
  pages     = {235105},
  volume    = {85},
  doi       = {10.1103/physrevb.85.235105},
  fjournal  = {Physical Review B},
  publisher = {American Physical Society (APS)},
}

@Article{Krausz2009,
  author    = {Krausz, Ferenc and Ivanov, Misha},
  journal   = {Rev. Mod. Phys.},
  title     = {Attosecond physics},
  year      = {2009},
  issn      = {1539-0756},
  month     = feb,
  number    = {1},
  pages     = {163--234},
  volume    = {81},
  doi       = {10.1103/revmodphys.81.163},
  fjournal  = {Reviews of Modern Physics},
  publisher = {American Physical Society (APS)},
}

@Article{List2015,
  author    = {List, Nanna Holmgaard and Kauczor, Joanna and Saue, Trond and Jensen, Hans Jørgen Aagaard and Norman, Patrick},
  journal   = {J. Chem. Phys.},
  title     = {Beyond the electric-dipole approximation: A formulation and implementation of molecular response theory for the description of absorption of electromagnetic field radiation},
  year      = {2015},
  issn      = {1089-7690},
  month     = jun,
  number    = {24},
  volume    = {142},
  doi       = {10.1063/1.4922697},
  fjournal  = {The Journal of Chemical Physics},
  publisher = {AIP Publishing},
}

@Book{Nielsen2012,
  author    = {Nielsen, Michael A. and Chuang, Isaac L.},
  publisher = {Cambridge University Press},
  title     = {Quantum computation and quantum information},
  year      = {2012},
  edition   = {10th anniversary ed., repr.},
  isbn      = {9781107002173},
  pagetotal = {676},
  ppn_gvk   = {1618949691},
}

@Article{Mishra2012,
  author    = {Mishra, R. and Kalita, D.J. and Gupta, A.K.},
  journal   = {Eur. Phys. J. D},
  title     = {Breakdown of dipole approximation and its effect on high harmonic generation},
  year      = {2012},
  issn      = {1434-6079},
  month     = jun,
  number    = {6},
  volume    = {66},
  doi       = {10.1140/epjd/e2012-20750-0},
  fjournal  = {The European Physical Journal D},
  publisher = {Springer Science and Business Media LLC},
}

@Article{Lindle1999,
  author    = {Lindle, Dennis W. and Hemmers, Oliver},
  journal   = {J. Electron Spectrosc.},
  title     = {Breakdown of the dipole approximation in soft-X-ray photoemission},
  year      = {1999},
  issn      = {0368-2048},
  month     = oct,
  number    = {1–3},
  pages     = {297--311},
  volume    = {100},
  doi       = {10.1016/s0368-2048(99)00052-3},
  fjournal  = {Journal of Electron Spectroscopy and Related Phenomena},
  publisher = {Elsevier BV},
}

@Article{Ludwig2014,
  author    = {Ludwig, A. and Maurer, J. and Mayer, B.W. and Phillips, C.R. and Gallmann, L. and Keller, U.},
  journal   = {Phys. Rev. Lett.},
  title     = {Breakdown of the Dipole Approximation in Strong-Field Ionization},
  year      = {2014},
  issn      = {1079-7114},
  month     = dec,
  number    = {24},
  pages     = {243001},
  volume    = {113},
  doi       = {10.1103/physrevlett.113.243001},
  fjournal  = {Physical Review Letters},
  publisher = {American Physical Society (APS)},
}

@Article{Torre2021,
  author    = {de la Torre, Alberto and Kennes, Dante M. and Claassen, Martin and Gerber, Simon and McIver, James W. and Sentef, Michael A.},
  journal   = {Rev. Mod. Phys.},
  title     = {Colloquium: Nonthermal pathways to ultrafast control in quantum materials},
  year      = {2021},
  issn      = {1539-0756},
  month     = oct,
  number    = {4},
  pages     = {041002},
  volume    = {93},
  doi       = {10.1103/revmodphys.93.041002},
  fjournal  = {Reviews of Modern Physics},
  publisher = {American Physical Society (APS)},
}

@Article{Kruchinin2018,
  author    = {Kruchinin, Stanislav Yu. and Krausz, Ferenc and Yakovlev, Vladislav S.},
  journal   = {Rev. Mod. Phys.},
  title     = {Colloquium : Strong-field phenomena in periodic systems},
  year      = {2018},
  issn      = {1539-0756},
  month     = apr,
  number    = {2},
  pages     = {021002},
  volume    = {90},
  doi       = {10.1103/revmodphys.90.021002},
  fjournal  = {Reviews of Modern Physics},
  publisher = {American Physical Society (APS)},
}

@Article{Schultze2012,
  author    = {Schultze, Martin and Bothschafter, Elisabeth M. and Sommer, Annkatrin and Holzner, Simon and Schweinberger, Wolfgang and Fiess, Markus and Hofstetter, Michael and Kienberger, Reinhard and Apalkov, Vadym and Yakovlev, Vladislav S. and Stockman, Mark I. and Krausz, Ferenc},
  journal   = {Nature},
  title     = {Controlling dielectrics with the electric field of light},
  year      = {2012},
  issn      = {1476-4687},
  month     = dec,
  number    = {7430},
  pages     = {75--78},
  volume    = {493},
  doi       = {10.1038/nature11720},
  publisher = {Springer Science and Business Media LLC},
}

@Article{Power1959,
  author    = {Power},
  journal   = {Philosophical Transactions of the Royal Society of London. Series A, Mathematical and Physical Sciences},
  title     = {Coulomb gauge in non-relativistic quantum electro-dynamics and the shape of spectral lines},
  year      = {1959},
  issn      = {2054-0272},
  month     = sep,
  number    = {999},
  pages     = {427--454},
  volume    = {251},
  doi       = {10.1098/rsta.1959.0008},
  publisher = {The Royal Society},
}

@Article{Noda2019,
  author    = {Noda, Masashi and Iida, Kenji and Yamaguchi, Maiku and Yatsui, Takashi and Nobusada, Katsuyuki},
  journal   = {Phys. Rev. Applied},
  title     = {Direct Wave-Vector Excitation in an Indirect-Band-Gap Semiconductor of Silicon with an Optical Near-field},
  year      = {2019},
  issn      = {2331-7019},
  month     = apr,
  number    = {4},
  pages     = {044053},
  volume    = {11},
  doi       = {10.1103/physrevapplied.11.044053},
  fjournal  = {Physical Review Applied},
  publisher = {American Physical Society (APS)},
}

@Article{Schueler2021,
  author    = {Schüler, Michael and Marks, Jacob A. and Murakami, Yuta and Jia, Chunjing and Devereaux, Thomas P.},
  journal   = {Phys. Rev. B},
  title     = {Gauge invariance of light-matter interactions in first-principle tight-binding models},
  year      = {2021},
  issn      = {2469-9969},
  month     = apr,
  number    = {15},
  pages     = {155409},
  volume    = {103},
  doi       = {10.1103/physrevb.103.155409},
  fjournal  = {Physical Review B},
  publisher = {American Physical Society (APS)},
}

@Article{Horn2022,
  author    = {van Horn, Martin and Saue, Trond and List, Nanna Holmgaard},
  journal   = {The Journal of Chemical Physics},
  title     = {Probing chirality across the electromagnetic spectrum with the full semi-classical light–matter interaction},
  year      = {2022},
  issn      = {1089-7690},
  month     = feb,
  number    = {5},
  volume    = {156},
  doi       = {10.1063/5.0077502},
  publisher = {AIP Publishing},
}

@Article{Golub1969,
  author    = {Golub, Gene H. and Welsch, John H.},
  journal   = {Mathematics of Computation},
  title     = {Calculation of Gauss quadrature rules},
  year      = {1969},
  issn      = {1088-6842},
  number    = {106},
  pages     = {221--230},
  volume    = {23},
  doi       = {10.1090/s0025-5718-69-99647-1},
  publisher = {American Mathematical Society (AMS)},
}

@Article{Bernadotte2012,
  author    = {Bernadotte, Stephan and Atkins, Andrew J. and Jacob, Christoph R.},
  journal   = {The Journal of Chemical Physics},
  title     = {Origin-independent calculation of quadrupole intensities in X-ray spectroscopy},
  year      = {2012},
  issn      = {1089-7690},
  month     = nov,
  number    = {20},
  volume    = {137},
  doi       = {10.1063/1.4766359},
  publisher = {AIP Publishing},
}

@Article{Hale2013,
  author    = {Hale, Nicholas and Townsend, Alex},
  journal   = {SIAM Journal on Scientific Computing},
  title     = {Fast and Accurate Computation of Gauss--Legendre and Gauss--Jacobi Quadrature Nodes and Weights},
  year      = {2013},
  issn      = {1095-7197},
  month     = jan,
  number    = {2},
  pages     = {A652--A674},
  volume    = {35},
  doi       = {10.1137/120889873},
  publisher = {Society for Industrial & Applied Mathematics (SIAM)},
}

@Article{Tiwari2025a,
  author    = {Tiwari, Vishal and Korol, Roman and Franco, Ignacio},
  journal   = {Physical Review Letters},
  title     = {Robust Purely Optical Signatures of Floquet States in Laser-Dressed Crystals},
  year      = {2025},
  issn      = {1079-7114},
  month     = oct,
  number    = {18},
  volume    = {135},
  doi       = {10.1103/5ywx-7dbs},
  publisher = {American Physical Society (APS)},
}

@Article{Oka2019,
  author    = {Oka, Takashi and Kitamura, Sota},
  journal   = {Annual Review of Condensed Matter Physics},
  title     = {Floquet Engineering of Quantum Materials},
  year      = {2019},
  issn      = {1947-5462},
  month     = mar,
  number    = {1},
  pages     = {387--408},
  volume    = {10},
  doi       = {10.1146/annurev-conmatphys-031218-013423},
  publisher = {Annual Reviews},
}

@Article{Merboldt2025,
  author    = {Merboldt, Marco and Schüler, Michael and Schmitt, David and Bange, Jan Philipp and Bennecke, Wiebke and Gadge, Karun and Pierz, Klaus and Schumacher, Hans Werner and Momeni, Davood and Steil, Daniel and Manmana, Salvatore R. and Sentef, Michael A. and Reutzel, Marcel and Mathias, Stefan},
  journal   = {Nature Physics},
  title     = {Observation of Floquet states in graphene},
  year      = {2025},
  issn      = {1745-2481},
  month     = may,
  number    = {7},
  pages     = {1093--1099},
  volume    = {21},
  doi       = {10.1038/s41567-025-02889-7},
  publisher = {Springer Science and Business Media LLC},
}

@Article{Ito2023,
  author    = {Ito, S. and Schüler, M. and Meierhofer, M. and Schlauderer, S. and Freudenstein, J. and Reimann, J. and Afanasiev, D. and Kokh, K. A. and Tereshchenko, O. E. and Güdde, J. and Sentef, M. A. and Höfer, U. and Huber, R.},
  journal   = {Nature},
  title     = {Build-up and dephasing of Floquet–Bloch bands on subcycle timescales},
  year      = {2023},
  issn      = {1476-4687},
  month     = apr,
  number    = {7958},
  pages     = {696--701},
  volume    = {616},
  doi       = {10.1038/s41586-023-05850-x},
  publisher = {Springer Science and Business Media LLC},
}

@Article{Schoetz2019,
  author    = {Schoetz, J. and Wang, Z. and Pisanty, E. and Lewenstein, M. and Kling, M. F. and Ciappina, M. F.},
  journal   = {ACS Photonics},
  title     = {Perspective on Petahertz Electronics and Attosecond Nanoscopy},
  year      = {2019},
  issn      = {2330-4022},
  month     = nov,
  number    = {12},
  pages     = {3057--3069},
  volume    = {6},
  doi       = {10.1021/acsphotonics.9b01188},
  publisher = {American Chemical Society (ACS)},
}

@Article{Hassan2024,
  author    = {Hassan, Mohammed Th.},
  journal   = {ACS Photonics},
  title     = {Lightwave Electronics: Attosecond Optical Switching},
  year      = {2024},
  issn      = {2330-4022},
  month     = feb,
  number    = {2},
  pages     = {334--338},
  volume    = {11},
  doi       = {10.1021/acsphotonics.3c01584},
  publisher = {American Chemical Society (ACS)},
}

@Article{Ji2019,
  author    = {Ji, Zhurun and Liu, Gerui and Addison, Zachariah and Liu, Wenjing and Yu, Peng and Gao, Heng and Liu, Zheng and Rappe, Andrew M. and Kane, Charles L. and Mele, Eugene J. and Agarwal, Ritesh},
  journal   = {Nature Materials},
  title     = {Spatially dispersive circular photogalvanic effect in a Weyl semimetal},
  year      = {2019},
  issn      = {1476-4660},
  month     = July,
  number    = {9},
  pages     = {955--962},
  volume    = {18},
  doi       = {10.1038/s41563-019-0421-5},
  publisher = {Springer Science and Business Media LLC},
}

@misc{supmat,
  note = {See Supplemental Material at [url] for additional details on the theoretical
  formulation, charge neutrality, convergence
  tests, higher-order multipole expansions,
  and computational performance.}
}

@misc{figshare_data,
  title = {Data for Light-Matter Interactions Beyond the Dipole Approximation in Extended Systems Without Multipole Expansion},
  year = {2026},
  howpublished = {Figshare},
  doi = {10.6084/m9.figshare.34895796},
  note = {Available at \url{https://doi.org/10.6084/m9.figshare.34895796}}
}

@Article{Ji2024,
  author    = {Ji, Zhurun and Zhao, Yuzhou and Chen, Yicong and Zhu, Ziyan and Wang, Yuhui and Liu, Wenjing and Modi, Gaurav and Mele, Eugene J. and Jin, Song and Agarwal, Ritesh},
  journal   = {Nature},
  title     = {Opto-twistronic Hall effect in a three-dimensional spiral lattice},
  year      = {2024},
  issn      = {1476-4687},
  month     = Sept,
  number    = {8032},
  pages     = {69--73},
  volume    = {634},
  doi       = {10.1038/s41586-024-07949-1},
  publisher = {Springer Science and Business Media LLC},
}

@Article{Beaulieu2024,
  author    = {Beaulieu, Samuel and Dong, Shuo and Christiansson, Viktor and Werner, Philipp and Pincelli, Tommaso and Ziegler, Jonas D. and Taniguchi, Takashi and Watanabe, Kenji and Chernikov, Alexey and Wolf, Martin and Rettig, Laurenz and Ernstorfer, Ralph and Schüler, Michael},
  journal   = {Science Advances},
  title     = {Berry curvature signatures in chiroptical excitonic transitions},
  year      = {2024},
  issn      = {2375-2548},
  month     = jun,
  number    = {26},
  volume    = {10},
  doi       = {10.1126/sciadv.adk3897},
  publisher = {American Association for the Advancement of Science (AAAS)},
}

@Article{Zhang2022,
  author    = {Zhang, Hongyun and Pincelli, Tommaso and Jozwiak, Chris and Kondo, Takeshi and Ernstorfer, Ralph and Sato, Takafumi and Zhou, Shuyun},
  journal   = {Nature Reviews Methods Primers},
  title     = {Angle-resolved photoemission spectroscopy},
  year      = {2022},
  issn      = {2662-8449},
  month     = jul,
  number    = {1},
  volume    = {2},
  doi       = {10.1038/s43586-022-00133-7},
  publisher = {Springer Science and Business Media LLC},
}

@Article{Babiker1983,
  author    = {Babiker, M. and Loudon, Rodney},
  journal   = {Proceedings of the Royal Society of London. A. Mathematical and Physical Sciences},
  title     = {Derivation of the Power-Zienau-Woolley Hamiltonian in quantum electrodynamics by gauge transformation},
  year      = {1983},
  issn      = {2053-9169},
  month     = Feb,
  number    = {1789},
  pages     = {439--460},
  volume    = {385},
  doi       = {10.1098/rspa.1983.0022},
  publisher = {The Royal Society},
}

@Article{UzanNarovlansky2024,
  author    = {Uzan-Narovlansky, Ayelet J. and Faeyrman, Lior and Brown, Graham G. and Shames, Sergei and Narovlansky, Vladimir and Xiao, Jiewen and Arusi-Parpar, Talya and Kneller, Omer and Bruner, Barry D. and Smirnova, Olga and Silva, Rui E. F. and Yan, Binghai and Jim\'enez-Gal\'an, \'Alvaro and Ivanov, Misha and Dudovich, Nirit},
  journal   = {Nature},
  title     = {Observation of interband Berry phase in laser-driven crystals},
  year      = {2024},
  issn      = {1476-4687},
  month     = jan,
  number    = {7997},
  pages     = {66--71},
  volume    = {626},
  doi       = {10.1038/s41586-023-06828-5},
  publisher = {Springer Science and Business Media LLC},
}

@Article{Zhang2025,
  author    = {Zhang, Zhuquan and Chien, Yu-Che and Wong, Man Tou and Gao, Frank Y. and Liu, Zi-Jie and Ma, Xiaoxuan and Cao, Shixun and Baldini, Edoardo and Nelson, Keith A.},
  journal   = {Science Advances},
  title     = {Terahertz stimulated parametric downconversion of a magnon mode in an antiferromagnet},
  year      = {2025},
  issn      = {2375-2548},
  month     = may,
  number    = {21},
  volume    = {11},
  doi       = {10.1126/sciadv.adv3757},
  publisher = {American Association for the Advancement of Science (AAAS)},
}

@Article{Zhang2024,
  author    = {Zhang, Zhuquan and Gao, Frank Y. and Curtis, Jonathan B. and Liu, Zi-Jie and Chien, Yu-Che and von Hoegen, Alexander and Wong, Man Tou and Kurihara, Takayuki and Suemoto, Tohru and Narang, Prineha and Baldini, Edoardo and Nelson, Keith A.},
  journal   = {Nature Physics},
  title     = {Terahertz field-induced nonlinear coupling of two magnon modes in an antiferromagnet},
  year      = {2024},
  issn      = {1745-2481},
  month     = jan,
  number    = {5},
  pages     = {801--806},
  volume    = {20},
  doi       = {10.1038/s41567-024-02386-3},
  publisher = {Springer Science and Business Media LLC},
}

@Article{McIver2019,
  author    = {McIver, J. W. and Schulte, B. and Stein, F.-U. and Matsuyama, T. and Jotzu, G. and Meier, G. and Cavalleri, A.},
  journal   = {Nature Physics},
  title     = {Light-induced anomalous Hall effect in graphene},
  year      = {2019},
  issn      = {1745-2481},
  month     = nov,
  number    = {1},
  pages     = {38--41},
  volume    = {16},
  doi       = {10.1038/s41567-019-0698-y},
  publisher = {Springer Science and Business Media LLC},
}

@Article{Keren2026,
  author    = {Keren, Itai and Webb, Tatiana A. and Zhang, Shuai and Xu, Jikai and Sun, Dihao and Kim, Brian S. Y. and Shin, Dongbin and Zhang, Songtian S. and Zhang, Junhe and Pereira, Giancarlo and Yao, Juntao and Okugawa, Takuya and Michael, Marios H. and Viñas Boström, Emil and Edgar, James H. and Wolf, Stuart and Julian, Matthew and Prasankumar, Rohit P. and Miyagawa, Kazuya and Kanoda, Kazushi and Gu, Genda and Cothrine, Matthew and Mandrus, David and Buzzi, Michele and Cavalleri, Andrea and Dean, Cory R. and Kennes, Dante M. and Millis, Andrew J. and Li, Qiang and Sentef, Michael A. and Rubio, Angel and Pasupathy, Abhay N. and Basov, D. N.},
  journal   = {Nature},
  title     = {Cavity-altered superconductivity},
  year      = {2026},
  issn      = {1476-4687},
  month     = feb,
  number    = {8103},
  pages     = {864--868},
  volume    = {650},
  doi       = {10.1038/s41586-025-10062-6},
  publisher = {Springer Science and Business Media LLC},
}

@Article{Zhou2023,
  author    = {Zhou, Shaohua and Bao, Changhua and Fan, Benshu and Wang, Fei and Zhong, Haoyuan and Zhang, Hongyun and Tang, Peizhe and Duan, Wenhui and Zhou, Shuyun},
  journal   = {Physical Review Letters},
  title     = {Floquet Engineering of Black Phosphorus upon Below-Gap Pumping},
  year      = {2023},
  issn      = {1079-7114},
  month     = sep,
  number    = {11},
  pages     = {116401},
  volume    = {131},
  doi       = {10.1103/physrevlett.131.116401},
  publisher = {American Physical Society (APS)},
}

@Article{Tiwari2025b,
  author    = {Tiwari, Vishal and Sierra-Ossa, Luis and Wojcik, Pawel and Franco, Ignacio},
  journal   = {The Journal of Physical Chemistry Letters},
  title     = {Transition from the Nanoscale to Bulk in the Nonequilibrium Optical Response of Laser-Dressed Materials},
  year      = {2025},
  issn      = {1948-7185},
  month     = dec,
  number    = {1},
  pages     = {214--221},
  volume    = {17},
  doi       = {10.1021/acs.jpclett.5c02710},
  publisher = {American Chemical Society (ACS)},
}

@Article{Tiwari2025,
  author    = {Tiwari, Vishal and Franco, Ignacio},
  journal   = {Physical Review B},
  title     = {First-principle-based Floquet engineering of solids in the velocity gauge},
  year      = {2025},
  issn      = {2469-9969},
  month     = aug,
  number    = {8},
  volume    = {112},
  doi       = {10.1103/2k9g-r77f},
  publisher = {American Physical Society (APS)},
}

@Article{Sennary2025,
  author    = {Sennary, Mohamed and Shah, Jalil and Yuan, Mingrui and Mahjoub, Ahmed and Pervak, Vladimir and Golubev, Nikolay V. and Hassan, Mohammed Th.},
  journal   = {Nature Communications},
  title     = {Light-induced quantum tunnelling current in graphene},
  year      = {2025},
  issn      = {2041-1723},
  month     = may,
  number    = {1},
  volume    = {16},
  doi       = {10.1038/s41467-025-59675-5},
  publisher = {Springer Science and Business Media LLC},
}

@Book{Shapiro2012,
  author    = {Shapiro, Moshe},
  editor    = {Paul Brumer},
  publisher = {John Wiley \& Sons, Incorporated},
  title     = {Quantum Control of Molecular Processes},
  year      = {2012},
  edition   = {2nd ed.},
  isbn      = {9783527639717},
  pagetotal = {1564},
  ppn_gvk   = {174642530X},
}

@Article{LHuillier2024,
  author    = {L’Huillier, Anne},
  journal   = {Reviews of Modern Physics},
  title     = {Nobel Lecture: The route to attosecond pulses},
  year      = {2024},
  issn      = {1539-0756},
  month     = aug,
  number    = {3},
  pages     = {030503},
  volume    = {96},
  doi       = {10.1103/revmodphys.96.030503},
  publisher = {American Physical Society (APS)},
}

@Article{Kim2022,
  author    = {Kim, San and Jeong, Tae-In and Park, Jongkyoon and Ciappina, Marcelo F. and Kim, Seungchul},
  journal   = {Nanophotonics},
  title     = {Recent advances in ultrafast plasmonics: from strong field physics to ultraprecision spectroscopy},
  year      = {2022},
  issn      = {2192-8614},
  month     = mar,
  number    = {11},
  pages     = {2393--2431},
  volume    = {11},
  doi       = {10.1515/nanoph-2021-0694},
  publisher = {Walter de Gruyter GmbH},
}

@Article{Xiao2017,
  author    = {Xiao, Lifu and Schultz, Zachary D.},
  journal   = {Analytical Chemistry},
  title     = {Spectroscopic Imaging at the Nanoscale: Technologies and Recent Applications},
  year      = {2017},
  issn      = {1520-6882},
  month     = oct,
  number    = {1},
  pages     = {440--458},
  volume    = {90},
  doi       = {10.1021/acs.analchem.7b04151},
  publisher = {American Chemical Society (ACS)},
}

@Article{Ciappina2017,
  author    = {Ciappina, M F and Pérez-Hernández, J A and Landsman, A S and Okell, W A and Zherebtsov, S and Förg, B and Schötz, J and Seiffert, L and Fennel, T and Shaaran, T and Zimmermann, T and Chacón, A and Guichard, R and Zaïr, A and Tisch, J W G and Marangos, J P and Witting, T and Braun, A and Maier, S A and Roso, L and Krüger, M and Hommelhoff, P and Kling, M F and Krausz, F and Lewenstein, M},
  journal   = {Reports on Progress in Physics},
  title     = {Attosecond physics at the nanoscale},
  year      = {2017},
  issn      = {1361-6633},
  month     = mar,
  number    = {5},
  pages     = {054401},
  volume    = {80},
  doi       = {10.1088/1361-6633/aa574e},
  publisher = {IOP Publishing},
}

@Article{Goulielmakis2022,
  author    = {Goulielmakis, Eleftherios and Brabec, Thomas},
  journal   = {Nature Photonics},
  title     = {High harmonic generation in condensed matter},
  year      = {2022},
  issn      = {1749-4893},
  month     = may,
  number    = {6},
  pages     = {411--421},
  volume    = {16},
  doi       = {10.1038/s41566-022-00988-y},
  publisher = {Springer Science and Business Media LLC},
}

@Article{Cavaletto2024,
  author    = {Cavaletto, Stefano M. and Kowalczyk, Katarzyna M. and Navarrete, Francisco O. and Rivera-Dean, Javier},
  journal   = {Nature Reviews Physics},
  title     = {The attoscience of strong-field-driven solids},
  year      = {2024},
  issn      = {2522-5820},
  month     = dec,
  number    = {1},
  pages     = {38--49},
  volume    = {7},
  doi       = {10.1038/s42254-024-00784-3},
  publisher = {Springer Science and Business Media LLC},
}

@Book{Milonni2010,
  author    = {Milonni, Peter W. and Eberly, Joseph H.},
  publisher = {Wiley},
  title     = {Laser Physics},
  year      = {2010},
  isbn      = {9780470409718},
  month     = mar,
  doi       = {10.1002/9780470409718},
}

@Article{Ghimire2018,
  author    = {Ghimire, Shambhu and Reis, David A.},
  journal   = {Nature Physics},
  title     = {High-harmonic generation from solids},
  year      = {2018},
  issn      = {1745-2481},
  month     = nov,
  number    = {1},
  pages     = {10--16},
  volume    = {15},
  doi       = {10.1038/s41567-018-0315-5},
  publisher = {Springer Science and Business Media LLC},
}

@Article{Dombi2020,
  author    = {Dombi, Péter and Pápa, Zsuzsanna and Vogelsang, Jan and Yalunin, Sergey V. and Sivis, Murat and Herink, Georg and Schäfer, Sascha and Groß, Petra and Ropers, Claus and Lienau, Christoph},
  journal   = {Reviews of Modern Physics},
  title     = {Strong-field nano-optics},
  year      = {2020},
  issn      = {1539-0756},
  month     = jun,
  number    = {2},
  pages     = {025003},
  volume    = {92},
  doi       = {10.1103/revmodphys.92.025003},
  publisher = {American Physical Society (APS)},
}

@Book{Craig2012,
  author    = {Craig, David P.},
  editor    = {T. Thirunamachandran},
  publisher = {Dover Publications},
  title     = {Molecular Quantum Electrodynamics},
  year      = {2012},
  isbn      = {9780486402147},
  series    = {Dover Books on Chemistry},
  pagetotal = {595},
  ppn_gvk   = {813277078},
}

@Book{Haug2008,
  editor    = {Hartmut Haug and Antti-Pekka Jauho},
  publisher = {Springer-Verlag Berlin Heidelberg},
  title     = {Quantum Kinetics in Transport and Optics of Semiconductors},
  year      = {2008},
  isbn      = {9783540735649},
  series    = {SpringerLink},
  ppn_gvk   = {1646421493},
}

@Book{Mandel1995,
  author    = {Mandel, Leonard and Wolf, Emil},
  publisher = {Cambridge University Press},
  title     = {Optical Coherence and Quantum Optics},
  year      = {1995},
  isbn      = {9781139644105},
  month     = sep,
  doi       = {10.1017/cbo9781139644105},
}

@Article{Moerner2002,
  author    = {Moerner, W. E.},
  journal   = {The Journal of Physical Chemistry B},
  title     = {A Dozen Years of Single-Molecule Spectroscopy in Physics, Chemistry, and Biophysics},
  year      = {2002},
  issn      = {1520-5207},
  month     = jan,
  number    = {5},
  pages     = {910--927},
  volume    = {106},
  doi       = {10.1021/jp012992g},
  publisher = {American Chemical Society (ACS)},
}

@Article{Cho2008,
  author    = {Cho, Minhaeng},
  journal   = {Chemical Reviews},
  title     = {Coherent Two-Dimensional Optical Spectroscopy},
  year      = {2008},
  issn      = {1520-6890},
  month     = mar,
  number    = {4},
  pages     = {1331--1418},
  volume    = {108},
  doi       = {10.1021/cr078377b},
  publisher = {American Chemical Society (ACS)},
}

@Article{Biswas2022,
  author    = {Biswas, Somnath and Kim, JunWoo and Zhang, Xinzi and Scholes, Gregory D.},
  journal   = {Chemical Reviews},
  title     = {Coherent Two-Dimensional and Broadband Electronic Spectroscopies},
  year      = {2022},
  issn      = {1520-6890},
  month     = jan,
  number    = {3},
  pages     = {4257--4321},
  volume    = {122},
  doi       = {10.1021/acs.chemrev.1c00623},
  publisher = {American Chemical Society (ACS)},
}

@Article{Zong2023,
  author    = {Zong, Alfred and Nebgen, Bailey R. and Lin, Sheng-Chih and Spies, Jacob A. and Zuerch, Michael},
  journal   = {Nature Reviews Materials},
  title     = {Emerging ultrafast techniques for studying quantum materials},
  year      = {2023},
  issn      = {2058-8437},
  month     = feb,
  number    = {4},
  pages     = {224--240},
  volume    = {8},
  doi       = {10.1038/s41578-022-00530-0},
  publisher = {Springer Science and Business Media LLC},
}

@Article{Heide2024a,
  author    = {Heide, Christian and Kobayashi, Yuki and Haque, Sheikh Rubaiat Ul and Ghimire, Shambhu},
  journal   = {Nature Physics},
  title     = {Ultrafast high-harmonic spectroscopy of solids},
  year      = {2024},
  issn      = {1745-2481},
  month     = oct,
  number    = {10},
  pages     = {1546--1557},
  volume    = {20},
  doi       = {10.1038/s41567-024-02640-8},
  publisher = {Springer Science and Business Media LLC},
}

@Article{Boschini2024,
  author    = {Boschini, Fabio and Zonno, Marta and Damascelli, Andrea},
  journal   = {Reviews of Modern Physics},
  title     = {Time-resolved ARPES studies of quantum materials},
  year      = {2024},
  issn      = {1539-0756},
  month     = feb,
  number    = {1},
  pages     = {015003},
  volume    = {96},
  doi       = {10.1103/revmodphys.96.015003},
  publisher = {American Physical Society (APS)},
}

@Article{Kraus2018,
  author    = {Kraus, Peter M. and Zürch, Michael and Cushing, Scott K. and Neumark, Daniel M. and Leone, Stephen R.},
  journal   = {Nature Reviews Chemistry},
  title     = {The ultrafast X-ray spectroscopic revolution in chemical dynamics},
  year      = {2018},
  issn      = {2397-3358},
  month     = may,
  number    = {6},
  pages     = {82--94},
  volume    = {2},
  doi       = {10.1038/s41570-018-0008-8},
  publisher = {Springer Science and Business Media LLC},
}

@Article{Choi2025,
  author    = {Choi, Dongsung and Mogi, Masataka and De Giovannini, Umberto and Azoury, Doron and Lv, Baiqing and Su, Yifan and Hübener, Hannes and Rubio, Angel and Gedik, Nuh},
  journal   = {Nature Physics},
  title     = {Observation of Floquet–Bloch states in monolayer graphene},
  year      = {2025},
  issn      = {1745-2481},
  month     = may,
  number    = {7},
  pages     = {1100--1105},
  volume    = {21},
  doi       = {10.1038/s41567-025-02888-8},
  publisher = {Springer Science and Business Media LLC},
}

@Article{Hui2021,
  author    = {Hui, Dandan and Alqattan, Husain and Yamada, Shunsuke and Pervak, Vladimir and Yabana, Kazuhiro and Hassan, Mohammed Th.},
  journal   = {Nature Photonics},
  title     = {Attosecond electron motion control in dielectric},
  year      = {2021},
  issn      = {1749-4893},
  month     = dec,
  number    = {1},
  pages     = {33--37},
  volume    = {16},
  doi       = {10.1038/s41566-021-00918-4},
  publisher = {Springer Science and Business Media LLC},
}

@Article{Feher2025,
  author    = {Fehér, Beatrix and Hanus, Václav and Li, Weiwei and Pápa, Zsuzsanna and Budai, Judit and Paul, Pallabi and Szeghalmi, Adriana and Wang, Zilong and Kling, Matthias F. and Dombi, Péter},
  journal   = {Science Advances},
  title     = {Light field–controlled PHz currents in intrinsic metals},
  year      = {2025},
  issn      = {2375-2548},
  month     = jun,
  number    = {26},
  volume    = {11},
  doi       = {10.1126/sciadv.adv5406},
  publisher = {American Association for the Advancement of Science (AAAS)},
}

@Book{Abramowitz2013,
  editor    = {Milton Abramowitz and Irene A. Stegun},
  publisher = {Dover Publ.},
  title     = {Handbook of mathematical functions},
  year      = {2013},
  address   = {New York, NY},
  edition   = {9. Dover print.; [Nachdr. der Ausg. von 1972]},
  isbn      = {9780486612720},
  series    = {Dover books on mathematics},
  pagetotal = {1046},
  ppn_gvk   = {838971261},
  subtitle  = {With formulas, graphs, and mathematical tables},
}

@Article{Mohan2026,
  author    = {Mohan, Abhinay V. and Serrat, Carles},
  journal   = {The Journal of Physical Chemistry A},
  title     = {Two-Level Theory of Second-Order Nonlinear X-ray Response beyond the Electric-Dipole Approximation},
  year      = {2026},
  issn      = {1520-5215},
  month     = jan,
  number    = {4},
  pages     = {823--829},
  volume    = {130},
  doi       = {10.1021/acs.jpca.5c07630},
  publisher = {American Chemical Society (ACS)},
}

@Article{Nien2013,
  author    = {Nien, Li-Wei and Lin, Shih-Che and Chao, Bo-Kai and Chen, Miin-Jang and Li, Jia-Han and Hsueh, Chun-Hway},
  journal   = {The Journal of Physical Chemistry C},
  title     = {Giant Electric Field Enhancement and Localized Surface Plasmon Resonance by Optimizing Contour Bowtie Nanoantennas},
  year      = {2013},
  issn      = {1932-7455},
  month     = nov,
  number    = {47},
  pages     = {25004--25011},
  volume    = {117},
  doi       = {10.1021/jp408610q},
  publisher = {American Chemical Society (ACS)},
}

@Article{Ayuso2022,
  author    = {Ayuso, David and Ordonez, Andres F. and Smirnova, Olga},
  journal   = {Physical Chemistry Chemical Physics},
  title     = {Ultrafast chirality: the road to efficient chiral measurements},
  year      = {2022},
  issn      = {1463-9084},
  number    = {44},
  pages     = {26962--26991},
  volume    = {24},
  doi       = {10.1039/d2cp01009g},
  publisher = {Royal Society of Chemistry (RSC)},
}

@Article{Soerensen2019,
  author    = {Sørensen, Lasse Kragh and Kieri, Emil and Srivastav, Shruti and Lundberg, Marcus and Lindh, Roland},
  journal   = {Physical Review A},
  title     = {Implementation of a semiclassical light-matter interaction using the Gauss-Hermite quadrature: A simple alternative to the multipole expansion},
  year      = {2019},
  issn      = {2469-9934},
  month     = jan,
  number    = {1},
  pages     = {013419},
  volume    = {99},
  doi       = {10.1103/physreva.99.013419},
  publisher = {American Physical Society (APS)},
}

@Article{Horn2023,
  author    = {van Horn, Martin and List, Nanna Holmgaard and Saue, Trond},
  journal   = {The Journal of Chemical Physics},
  title     = {Transition moments beyond the electric-dipole approximation: Visualization and basis set requirements},
  year      = {2023},
  issn      = {1089-7690},
  month     = may,
  number    = {18},
  volume    = {158},
  doi       = {10.1063/5.0147105},
  publisher = {AIP Publishing},
}

@Article{Chernyak1995,
  author    = {Chernyak, Vladimir and Mukamel, Shaul},
  journal   = {Chem. Phys.},
  title     = {Gauge invariant formulation of molecular electrodynamics and the multipolar Hamiltonian},
  year      = {1995},
  issn      = {0301-0104},
  month     = sep,
  number    = {1–2},
  pages     = {133--143},
  volume    = {198},
  doi       = {10.1016/0301-0104(95)00122-5},
  fjournal  = {Chemical Physics},
  publisher = {Elsevier BV},
}

@Article{Iwasa2024,
  author    = {Iwasa, Takeshi},
  journal   = {The Journal of Physical Chemistry Letters},
  title     = {Generalized Transition Moment and Oscillator Strength for Optimal Control of Excited States Using Near-Field Light},
  year      = {2024},
  issn      = {1948-7185},
  month     = apr,
  number    = {18},
  pages     = {4775--4781},
  volume    = {15},
  doi       = {10.1021/acs.jpclett.4c00609},
  publisher = {American Chemical Society (ACS)},
}

@Article{Shan2021,
  author    = {Shan, Jun-Yi and Ye, M. and Chu, H. and Lee, Sungmin and Park, Je-Geun and Balents, L. and Hsieh, D.},
  journal   = {Nature},
  title     = {Giant modulation of optical nonlinearity by Floquet engineering},
  year      = {2021},
  issn      = {1476-4687},
  month     = dec,
  number    = {7888},
  pages     = {235--239},
  volume    = {600},
  doi       = {10.1038/s41586-021-04051-8},
  publisher = {Springer Science and Business Media LLC},
}

@Article{Giannozzi2009,
  author    = {Giannozzi, Paolo and Baroni, Stefano and Bonini, Nicola and Calandra, Matteo and Car, Roberto and Cavazzoni, Carlo and Ceresoli, Davide and Chiarotti, Guido L and Cococcioni, Matteo and Dabo, Ismaila and Dal Corso, Andrea and de Gironcoli, Stefano and Fabris, Stefano and Fratesi, Guido and Gebauer, Ralph and Gerstmann, Uwe and Gougoussis, Christos and Kokalj, Anton and Lazzeri, Michele and Martin-Samos, Layla and Marzari, Nicola and Mauri, Francesco and Mazzarello, Riccardo and Paolini, Stefano and Pasquarello, Alfredo and Paulatto, Lorenzo and Sbraccia, Carlo and Scandolo, Sandro and Sclauzero, Gabriele and Seitsonen, Ari P and Smogunov, Alexander and Umari, Paolo and Wentzcovitch, Renata M},
  journal   = {J. Phys. Condens. Matter},
  title     = {QUANTUM ESPRESSO: a modular and open-source software project for quantum simulations of materials},
  year      = {2009},
  issn      = {1361-648X},
  month     = sep,
  number    = {39},
  pages     = {395502},
  volume    = {21},
  doi       = {10.1088/0953-8984/21/39/395502},
  fjournal  = {Journal of Physics: Condensed Matter},
  publisher = {IOP Publishing},
}

@Article{Mostofi2014,
  author    = {Mostofi, Arash A. and Yates, Jonathan R. and Pizzi, Giovanni and Lee, Young-Su and Souza, Ivo and Vanderbilt, David and Marzari, Nicola},
  journal   = {Comput. Phys. Commun.},
  title     = {An updated version of wannier90: A tool for obtaining maximally-localised Wannier functions},
  year      = {2014},
  issn      = {0010-4655},
  month     = aug,
  number    = {8},
  pages     = {2309--2310},
  volume    = {185},
  doi       = {10.1016/j.cpc.2014.05.003},
  fjournal  = {Computer Physics Communications},
  publisher = {Elsevier BV},
}

@Book{Siegman1986,
  author    = {Siegman, Anthony E.},
  publisher = {University Science Books},
  title     = {Lasers},
  year      = {1986},
  address   = {Mill Valley, California},
  isbn      = {0935702113},
  pagetotal = {1283},
  ppn_gvk   = {1629380601},
}

@Article{Hindmarsh2005,
  author    = {Hindmarsh, Alan C. and Brown, Peter N. and Grant, Keith E. and Lee, Steven L. and Serban, Radu and Shumaker, Dan E. and Woodward, Carol S.},
  journal   = {ACM Transactions on Mathematical Software},
  title     = {SUNDIALS: Suite of nonlinear and differential/algebraic equation solvers},
  year      = {2005},
  issn      = {1557-7295},
  month     = sep,
  number    = {3},
  pages     = {363--396},
  volume    = {31},
  doi       = {10.1145/1089014.1089020},
  publisher = {Association for Computing Machinery (ACM)},
}

@Article{Brennecke2018,
  author    = {Brennecke, Simon and Lein, Manfred},
  journal   = {J. Phys. B: At. Mol. Opt. Phys.},
  title     = {High-order above-threshold ionization beyond the electric dipole approximation},
  year      = {2018},
  issn      = {1361-6455},
  month     = apr,
  number    = {9},
  pages     = {094005},
  volume    = {51},
  doi       = {10.1088/1361-6455/aab91f},
  fjournal  = {Journal of Physics B: Atomic, Molecular and Optical Physics},
  publisher = {IOP Publishing},
}

@Article{Brabec2000,
  author    = {Brabec, Thomas and Krausz, Ferenc},
  journal   = {Rev. Mod. Phys.},
  title     = {Intense few-cycle laser fields: Frontiers of nonlinear optics},
  year      = {2000},
  issn      = {1539-0756},
  month     = apr,
  number    = {2},
  pages     = {545--591},
  volume    = {72},
  doi       = {10.1103/revmodphys.72.545},
  fjournal  = {Reviews of Modern Physics},
  publisher = {American Physical Society (APS)},
}

@Article{Yamaguchi2016a,
  author    = {Yamaguchi, Maiku and Nobusada, Katsuyuki},
  journal   = {The Journal of Physical Chemistry C},
  title     = {Large Hyperpolarizabilities of the Second Harmonic Generation Induced by Nonuniform Optical Near Fields},
  year      = {2016},
  issn      = {1932-7455},
  month     = oct,
  number    = {41},
  pages     = {23748--23755},
  volume    = {120},
  doi       = {10.1021/acs.jpcc.6b08507},
  publisher = {American Chemical Society (ACS)},
}

@Article{Franco2008,
  author    = {Franco, Ignacio and Shapiro, Moshe and Brumer, Paul},
  journal   = {J. Chem. Phys.},
  title     = {Laser-induced currents along molecular wire junctions},
  year      = {2008},
  issn      = {1089-7690},
  month     = jun,
  number    = {24},
  volume    = {128},
  doi       = {10.1063/1.2940796},
  fjournal  = {The Journal of Chemical Physics},
  publisher = {AIP Publishing},
}

@Article{Boolakee2022,
  author    = {Boolakee, Tobias and Heide, Christian and Garzón-Ramírez, Antonio and Weber, Heiko B. and Franco, Ignacio and Hommelhoff, Peter},
  journal   = {Nature},
  title     = {Light-field control of real and virtual charge carriers},
  year      = {2022},
  issn      = {1476-4687},
  month     = may,
  number    = {7909},
  pages     = {251--255},
  volume    = {605},
  doi       = {10.1038/s41586-022-04565-9},
  publisher = {Springer Science and Business Media LLC},
}

@Article{Higuchi2017,
  author    = {Higuchi, Takuya and Heide, Christian and Ullmann, Konrad and Weber, Heiko B. and Hommelhoff, Peter},
  journal   = {Nature},
  title     = {Light-field-driven currents in graphene},
  year      = {2017},
  issn      = {1476-4687},
  month     = sep,
  number    = {7675},
  pages     = {224--228},
  volume    = {550},
  doi       = {10.1038/nature23900},
  publisher = {Springer Science and Business Media LLC},
}

@Article{Woolley1971,
  author    = {R.G. WOOLEY},
  journal   = {Proceedings of the Royal Society of London. A. Mathematical and Physical Sciences},
  title     = {Molecular quantum electrodynamics},
  year      = {1971},
  issn      = {0080-4630},
  month     = mar,
  number    = {1547},
  pages     = {557--572},
  volume    = {321},
  doi       = {10.1098/rspa.1971.0049},
  publisher = {The Royal Society},
}

@Article{Bandrauk2006,
  author    = {Bandrauk, A. D. and Lu, H. Z.},
  journal   = {Phys. Rev. A},
  title     = {Molecules in intense laser fields: Beyond the dipole approximation},
  year      = {2006},
  issn      = {1094-1622},
  month     = jan,
  number    = {1},
  pages     = {013412},
  volume    = {73},
  doi       = {10.1103/physreva.73.013412},
  fjournal  = {Physical Review A},
  publisher = {American Physical Society (APS)},
}

@Article{Foerre2022,
  author    = {Førre, Morten},
  journal   = {Phys. Rev. A},
  title     = {Nondipole effects and photoelectron momentum shifts in strong-field ionization by infrared light},
  year      = {2022},
  issn      = {2469-9934},
  month     = jul,
  number    = {1},
  pages     = {013104},
  volume    = {106},
  doi       = {10.1103/physreva.106.013104},
  fjournal  = {Physical Review A},
  publisher = {American Physical Society (APS)},
}

@Book{Mukamel2009,
  author    = {Mukamel, Shaul},
  publisher = {Oxford Univ. Press},
  title     = {Principles of nonlinear optical spectroscopy},
  year      = {2009},
  isbn      = {9780195132915},
  number    = {6},
  series    = {Oxford series in optical and imaging sciences},
  pagetotal = {543},
  ppn_gvk   = {614912725},
}

@Article{Marzari2012,
  author    = {Marzari, Nicola and Mostofi, Arash A. and Yates, Jonathan R. and Souza, Ivo and Vanderbilt, David},
  journal   = {Rev. Mod. Phys.},
  title     = {Maximally localized Wannier functions: Theory and applications},
  year      = {2012},
  issn      = {1539-0756},
  month     = oct,
  number    = {4},
  pages     = {1419--1475},
  volume    = {84},
  doi       = {10.1103/revmodphys.84.1419},
  fjournal  = {Reviews of Modern Physics},
  publisher = {American Physical Society (APS)},
}

@Article{Foerre2006,
  author    = {Førre, M. and Hansen, J. P. and Kocbach, L. and Selstø, S. and Madsen, L. B.},
  journal   = {Phys. Rev. Lett.},
  title     = {Nondipole Ionization Dynamics of Atoms in Superintense High-Frequency Attosecond Pulses},
  year      = {2006},
  issn      = {1079-7114},
  month     = jul,
  number    = {4},
  pages     = {043601},
  volume    = {97},
  doi       = {10.1103/physrevlett.97.043601},
  fjournal  = {Physical Review Letters},
  publisher = {American Physical Society (APS)},
}

@Article{Suster2023,
  author    = {Suster, M. C. and Derlikiewicz, J. and Krajewska, K. and Vélez, F. Cajiao and Kamiński, J. Z.},
  journal   = {Phys. Rev. A},
  title     = {Nondipole signatures in ionization and high-order harmonic generation},
  year      = {2023},
  issn      = {2469-9934},
  month     = may,
  number    = {5},
  pages     = {053112},
  volume    = {107},
  doi       = {10.1103/physreva.107.053112},
  fjournal  = {Physical Review A},
  publisher = {American Physical Society (APS)},
}

@Article{Iwasa2009,
  author    = {Iwasa, Takeshi and Nobusada, Katsuyuki},
  journal   = {Phys. Rev. A},
  title     = {Nonuniform light-matter interaction theory for near-field-induced electron dynamics},
  year      = {2009},
  issn      = {1094-1622},
  month     = oct,
  number    = {4},
  pages     = {043409},
  volume    = {80},
  doi       = {10.1103/physreva.80.043409},
  fjournal  = {Physical Review A},
  publisher = {American Physical Society (APS)},
}

@Article{Ghimire2010,
  author    = {Ghimire, Shambhu and DiChiara, Anthony D. and Sistrunk, Emily and Agostini, Pierre and DiMauro, Louis F. and Reis, David A.},
  journal   = {Nat. Phys.},
  title     = {Observation of high-order harmonic generation in a bulk crystal},
  year      = {2010},
  issn      = {1745-2481},
  month     = dec,
  number    = {2},
  pages     = {138--141},
  volume    = {7},
  doi       = {10.1038/nphys1847},
  fjournal  = {Nature Physics},
  publisher = {Springer Science and Business Media LLC},
}

@Article{Demekhin2014,
  author    = {Demekhin, Ph V},
  journal   = {J. Phys. B: At. Mol. Opt. Phys.},
  title     = {On the breakdown of the electric dipole approximation for hard x-ray photoionization cross sections},
  year      = {2014},
  issn      = {1361-6455},
  month     = jan,
  number    = {2},
  pages     = {025602},
  volume    = {47},
  doi       = {10.1088/0953-4075/47/2/025602},
  fjournal  = {Journal of Physics B: Atomic, Molecular and Optical Physics},
  publisher = {IOP Publishing},
}

@Article{Woolley2020,
  author    = {Woolley, R. Guy},
  journal   = {Phys. Rev. Research},
  title     = {Power-Zienau-Woolley representations of nonrelativistic QED for atoms and molecules},
  year      = {2020},
  issn      = {2643-1564},
  month     = feb,
  number    = {1},
  pages     = {013206},
  volume    = {2},
  doi       = {10.1103/physrevresearch.2.013206},
  fjournal  = {Physical Review Research},
  publisher = {American Physical Society (APS)},
}

@Article{DalCorso2014,
  author    = {Dal Corso, Andrea},
  journal   = {Nato. Sc. S. Ss. Iii. C. S.},
  title     = {Pseudopotentials periodic table: From H to Pu},
  year      = {2014},
  issn      = {0927-0256},
  month     = dec,
  pages     = {337--350},
  volume    = {95},
  doi       = {10.1016/j.commatsci.2014.07.043},
  fjournal  = {Computational Materials Science},
  publisher = {Elsevier BV},
}

@Article{Franco2007,
  author    = {Franco, Ignacio and Shapiro, Moshe and Brumer, Paul},
  journal   = {Phys. Rev. Lett.},
  title     = {Robust Ultrafast Currents in Molecular Wires through Stark Shifts},
  year      = {2007},
  issn      = {1079-7114},
  month     = sep,
  number    = {12},
  pages     = {126802},
  volume    = {99},
  doi       = {10.1103/physrevlett.99.126802},
  fjournal  = {Physical Review Letters},
  publisher = {American Physical Society (APS)},
}

@Article{Noda2017,
  author    = {Noda, Masashi and Yamaguchi, Maiku and Nobusada, Katsuyuki},
  journal   = {The Journal of Physical Chemistry C},
  title     = {Second Harmonic Excitation of Acetylene by the Optical Near Field Generated in a Porous Material},
  year      = {2017},
  issn      = {1932-7455},
  month     = may,
  number    = {21},
  pages     = {11687--11692},
  volume    = {121},
  doi       = {10.1021/acs.jpcc.7b02744},
  publisher = {American Chemical Society (ACS)},
}

@Article{Perdew1981,
  author    = {Perdew, J. P. and Zunger, Alex},
  journal   = {Phys. Rev. B},
  title     = {Self-interaction correction to density-functional approximations for many-electron systems},
  year      = {1981},
  issn      = {0163-1829},
  month     = may,
  number    = {10},
  pages     = {5048--5079},
  volume    = {23},
  doi       = {10.1103/physrevb.23.5048},
  fjournal  = {Physical Review B},
  publisher = {American Physical Society (APS)},
}

@Article{Rybka2016,
  author    = {Rybka, Tobias and Ludwig, Markus and Schmalz, Michael F. and Knittel, Vanessa and Brida, Daniele and Leitenstorfer, Alfred},
  journal   = {Nat. Photonics},
  title     = {Sub-cycle optical phase control of nanotunnelling in the single-electron regime},
  year      = {2016},
  issn      = {1749-4893},
  month     = sep,
  number    = {10},
  pages     = {667--670},
  volume    = {10},
  doi       = {10.1038/nphoton.2016.174},
  fjournal  = {Nature Photonics},
  publisher = {Springer Science and Business Media LLC},
}

@Article{GarzonRamirez2020,
  author    = {Garzón-Ramírez, Antonio J. and Franco, Ignacio},
  journal   = {J. Chem. Phys.},
  title     = {Symmetry breaking in the Stark Control of Electrons at Interfaces (SCELI)},
  year      = {2020},
  issn      = {1089-7690},
  month     = jul,
  number    = {4},
  volume    = {153},
  doi       = {10.1063/5.0013190},
  fjournal  = {The Journal of Chemical Physics},
  publisher = {AIP Publishing},
}

@Book{CohenTannoudji1997,
  author    = {Cohen‐Tannoudji, Claude and Dupont‐Roc, Jacques and Grynberg, Gilbert},
  publisher = {Wiley},
  title     = {Photons and Atoms: Introduction to Quantum Electrodynamics},
  year      = {1997},
  isbn      = {9783527618422},
  month     = {mar},
  doi       = {10.1002/9783527618422},
}

@Article{Marzari1997,
  author    = {Marzari, Nicola and Vanderbilt, David},
  journal   = {Phys. Rev. B},
  title     = {Maximally localized generalized Wannier functions for composite energy bands},
  year      = {1997},
  issn      = {1095-3795},
  month     = nov,
  number    = {20},
  pages     = {12847--12865},
  volume    = {56},
  doi       = {10.1103/physrevb.56.12847},
  fjournal  = {Physical Review B},
  publisher = {American Physical Society (APS)},
}

@Article{Yamaguchi2015,
  author    = {Yamaguchi, Maiku and Nobusada, Katsuyuki and Kawazoe, Tadashi and Yatsui, Takashi},
  journal   = {Appl. Phys. Lett.},
  title     = {Two-photon absorption induced by electric field gradient of optical near-field and its application to photolithography},
  year      = {2015},
  issn      = {1077-3118},
  month     = may,
  number    = {19},
  volume    = {106},
  doi       = {10.1063/1.4921005},
  fjournal  = {Applied Physics Letters},
  publisher = {AIP Publishing},
}

@Article{Lestrange2015,
  author    = {Lestrange, Patrick J. and Egidi, Franco and Li, Xiaosong},
  journal   = {The Journal of Chemical Physics},
  title     = {The consequences of improperly describing oscillator strengths beyond the electric dipole approximation},
  year      = {2015},
  issn      = {1089-7690},
  month     = dec,
  number    = {23},
  volume    = {143},
  doi       = {10.1063/1.4937410},
  publisher = {AIP Publishing},
}

@Article{Aurbakken2024,
  author = {Aurbakken, Einar and Ofstad, Benedicte Sverdrup and Kristiansen, H{\aa}kon Emil and Sch{\o}yen, {\O}yvind Sigmundson and Kvaal, Simen and S{\o}rensen, Lasse Kragh and Lindh, Roland and Pedersen, Thomas Bondo},
  journal   = {Physical Review A},
  title     = {Transient spectroscopy from time-dependent electronic-structure theory without multipole expansions},
  year      = {2024},
  issn      = {2469-9934},
  month     = jan,
  number    = {1},
  pages     = {013109},
  volume    = {109},
  doi       = {10.1103/physreva.109.013109},
  publisher = {American Physical Society (APS)},
}

@Article{Jensen2022,
  author    = {Jensen, Simon Vendelbo Bylling and Madsen, Lars Bojer},
  journal   = {Physical Review A},
  title     = {Propagation time and nondipole contributions to intraband high-order harmonic generation},
  year      = {2022},
  issn      = {2469-9934},
  month     = feb,
  number    = {2},
  pages     = {l021101},
  volume    = {105},
  doi       = {10.1103/physreva.105.l021101},
  publisher = {American Physical Society (APS)},
}

@Article{Bonafe2025,
  author    = {Bonafé, Franco P. and Albar, Esra Ilke and Ohlmann, Sebastian T. and Kosheleva, Valeriia P. and Bustamante, Carlos M. and Troisi, Francesco and Rubio, Angel and Appel, Heiko},
  journal   = {Physical Review B},
  title     = {Full minimal coupling Maxwell-TDDFT: An ab initio framework for light-matter interaction beyond the dipole approximation},
  year      = {2025},
  issn      = {2469-9969},
  month     = feb,
  number    = {8},
  pages     = {085114},
  volume    = {111},
  doi       = {10.1103/physrevb.111.085114},
  publisher = {American Physical Society (APS)},
}

@InBook{Taflove2005,
  author    = {Taflove, Allen and Hagness, Susan C. and Piket-May, Melinda},
  pages     = {629--670},
  publisher = {Elsevier},
  title     = {Computational Electromagnetics: The Finite-Difference Time-Domain Method},
  year      = {2005},
  isbn      = {9780121709600},
  booktitle = {The Electrical Engineering Handbook},
  doi       = {10.1016/b978-012170960-0/50046-3},
}

@Article{Oskooi2010,
  author    = {Oskooi, Ardavan F. and Roundy, David and Ibanescu, Mihai and Bermel, Peter and Joannopoulos, J.D. and Johnson, Steven G.},
  journal   = {Computer Physics Communications},
  title     = {Meep: A flexible free-software package for electromagnetic simulations by the FDTD method},
  year      = {2010},
  issn      = {0010-4655},
  month     = mar,
  number    = {3},
  pages     = {687--702},
  volume    = {181},
  doi       = {10.1016/j.cpc.2009.11.008},
  publisher = {Elsevier BV},
}

@Article{Berenger1994,
  author    = {Berenger, Jean-Pierre},
  journal   = {Journal of Computational Physics},
  title     = {A perfectly matched layer for the absorption of electromagnetic waves},
  year      = {1994},
  issn      = {0021-9991},
  month     = oct,
  number    = {2},
  pages     = {185--200},
  volume    = {114},
  doi       = {10.1006/jcph.1994.1159},
  publisher = {Elsevier BV},
}

@Article{Khalil2021,
  author    = {Khalil, Usman Khan and Farooq, Waqas and Iqbal, Javed and Ul Abideen Kazmi, Syed Zain and Khan, Adnan Daud and Ur Rehman, Anees and Ayub, Saba},
  journal   = {The European Physical Journal Plus},
  title     = {Design and optimization of bowtie nanoantenna for electromagnetic field enhancement},
  year      = {2021},
  issn      = {2190-5444},
  month     = jul,
  number    = {7},
  volume    = {136},
  doi       = {10.1140/epjp/s13360-021-01702-7},
  publisher = {Springer Science and Business Media LLC},
}

@Article{Hofbrucker2020,
  author    = {Hofbrucker, J. and Volotka, A. V. and Fritzsche, S.},
  journal   = {Scientific Reports},
  title     = {Breakdown of the electric dipole approximation at Cooper minima in direct two-photon ionisation},
  year      = {2020},
  issn      = {2045-2322},
  month     = feb,
  number    = {1},
  volume    = {10},
  doi       = {10.1038/s41598-020-60206-z},
  publisher = {Springer Science and Business Media LLC},
}

@Article{Nishizawa2025,
  author    = {Nishizawa, Daisuke and Taketsugu, Tetsuya and Iwasa, Takeshi},
  journal   = {Chemistry Letters},
  title     = {Theoretical study of near-field-induced local excitation dynamics},
  year      = {2025},
  issn      = {1348-0715},
  month     = may,
  number    = {6},
  volume    = {54},
  doi       = {10.1093/chemle/upaf109},
  publisher = {Oxford University Press (OUP)},
}

@Book{Yu2010,
  author    = {Yu, Peter Y. and Cardona, Manuel},
  publisher = {Springer Berlin Heidelberg},
  title     = {Fundamentals of Semiconductors: Physics and Materials Properties},
  year      = {2010},
  isbn      = {9783642007101},
  doi       = {10.1007/978-3-642-00710-1},
  issn      = {1868-4521},
  journal   = {Graduate Texts in Physics},
}

@Article{Jensen2020,
  author    = {Jensen, Simon Vendelbo Bylling and Lund, Mads Middelhede and Madsen, Lars Bojer},
  journal   = {Physical Review A},
  title     = {Nondipole strong-field-approximation Hamiltonian},
  year      = {2020},
  issn      = {2469-9934},
  month     = apr,
  number    = {4},
  pages     = {043408},
  volume    = {101},
  doi       = {10.1103/physreva.101.043408},
  publisher = {American Physical Society (APS)},
}

@Article{DellaPicca2023,
  author    = {Della Picca, R. and Randazzo, J. M. and López, S. D. and Ciappina, M. F. and Arbó, D. G.},
  journal   = {Physical Review A},
  title     = {Laser-assisted photoionization beyond the dipole approximation},
  year      = {2023},
  issn      = {2469-9934},
  month     = may,
  number    = {5},
  pages     = {053104},
  volume    = {107},
  doi       = {10.1103/physreva.107.053104},
  publisher = {American Physical Society (APS)},
}

@Article{Jensen2025,
  author    = {Jensen, Simon Vendelbo Bylling and Tancogne-Dejean, Nicolas and Rubio, Angel and Madsen, Lars Bojer},
  journal   = {Physical Review Letters},
  title     = {Beyond Electric-Dipole Treatment of Light-Matter Interactions in Materials: Nondipole Harmonic Generation in Bulk Si},
  year      = {2025},
  issn      = {1079-7114},
  month     = may,
  number    = {19},
  pages     = {196902},
  volume    = {134},
  doi       = {10.1103/physrevlett.134.196902},
  publisher = {American Physical Society (APS)},
}

\begin{widetext}
\setcounter{equation}{0}
\setcounter{figure}{0}
\setcounter{section}{0}

\renewcommand{\theequation}{S\arabic{equation}}
\renewcommand{\thefigure}{S\arabic{figure}}
\renewcommand{\thepage}{S\arabic{page}}
\renewcommand{\thesection}{S\Roman{section}}

\section*{Supplementary Material}

\section{Inclusion of the inter-cell matrix elements}

As discussed in Sec. II B from the main text, the inter-cell contributions to the position matrix elements, $D_{m \mathbf{R}'n\mathbf{R}} = \langle m (\mathbf{R}'-\mathbf{R})| \hat{\mathbf{r}} |n \mathbf{0} \rangle $ for $\mathbf{R}'\neq \mathbf{R}$ are often negligible. In the main-text, we therefore retained only the intra-cell elements $D_{mn} = \langle m \mathbf{0}| \hat{\mathbf{r}} |n \mathbf{0} \rangle $, which  simplifies Eq. (8). 
Here, we explicitly assess the impact of including the inter-cell contributions. Figure~\ref{fig:intercell} compares the multipolar dynamics generated by Eq. (21) of the main text using only intra-cell matrix elements with those obtained when both intra- and inter-cell terms are retained. We consider the nonuniform illumination protocol introduced in Sec. III D1 of the main text.
The dynamics are indistinguishable in both cases, confirming the robustness of the \emph{modified} MLWFs strategy. For computational simplicity, we use Eq. (23) of the main text with $N_q=6$ when both intra- and inter-cell terms are taken into account, as including inter-cell contributions removes the block-diagonal structure of the interaction Hamiltonian [Eq. (21)].

\begin{figure*}[htb] 
\centering
\includegraphics[width=1.0\textwidth,keepaspectratio]{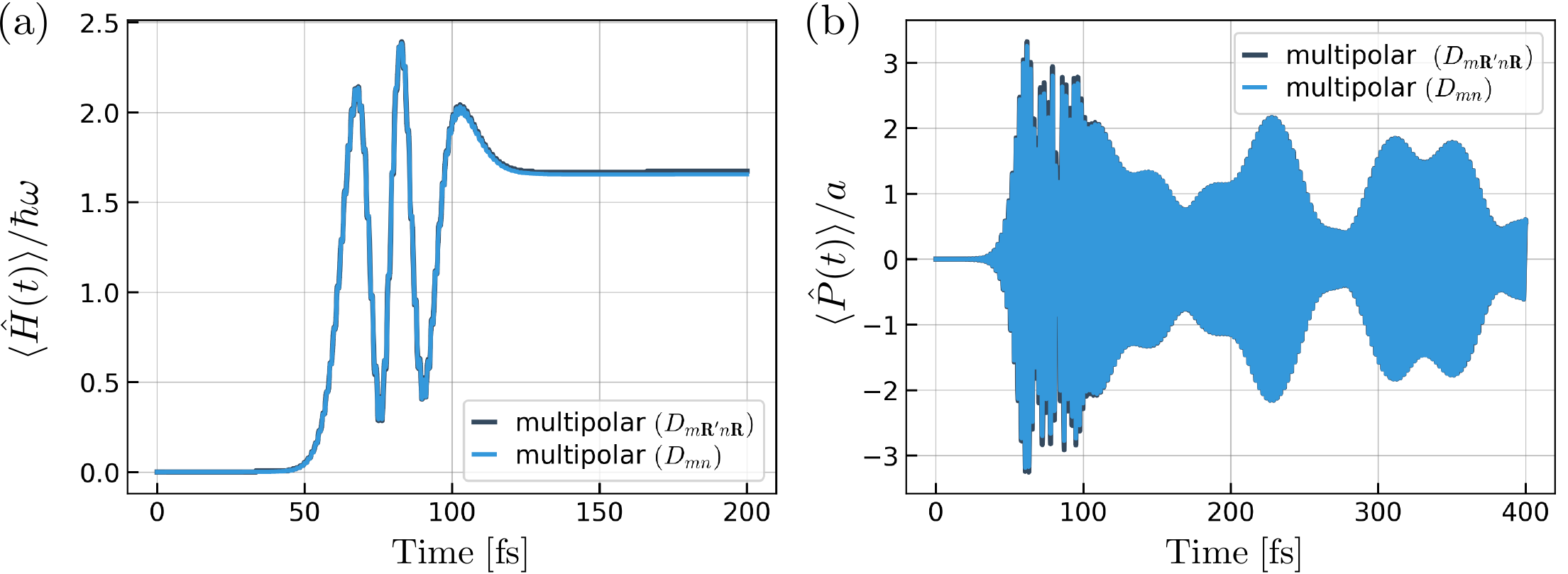}
\caption{The average energy (a) and polarization (b) of the chain with $N=80$ unit cells, comparing the dynamics generated by the multipolar PZW Hamiltonian with both inter-cell and intra-cell contribution included (black), to the dynamics generated with including only intra-cell contribution (blue). Simulation parameters are the same as in Figure 2 from the main text, except 0.2-$\mu$m-long chain.} 
\label{fig:intercell}
\end{figure*}

\section{Charge neutrality and origin independence} 

As described in Sec. III B of the main text, we subtract the reference electronic density from the light-matter interaction Hamiltonian so that the electric field couples to a charge-neutral system, thereby ensuring origin-independent dynamics.
To demonstrate this explicitly, we consider the nonuniform illumination of the chain introduced in Sec. III D1 of the main text and compare the multipolar dynamics generated by Eq. (21) of the main text, with the reference electronic density subtracted [Fig.~\ref{fig:origindependence}(b)] to those obtained without the subtraction [Fig.~\ref{fig:origindependence}(a)].
Upon shifting the coordinate origin by one unit-cell length $a$, we find that the dynamics generated by Eq. (21) remain invariant, confirming origin independence.
While the instantaneous average energy differs during the presence of the field, the net energy change after the pulse is identical in both cases.

\begin{figure*}[htb] 
\centering
\includegraphics[width=1.0\textwidth,keepaspectratio]{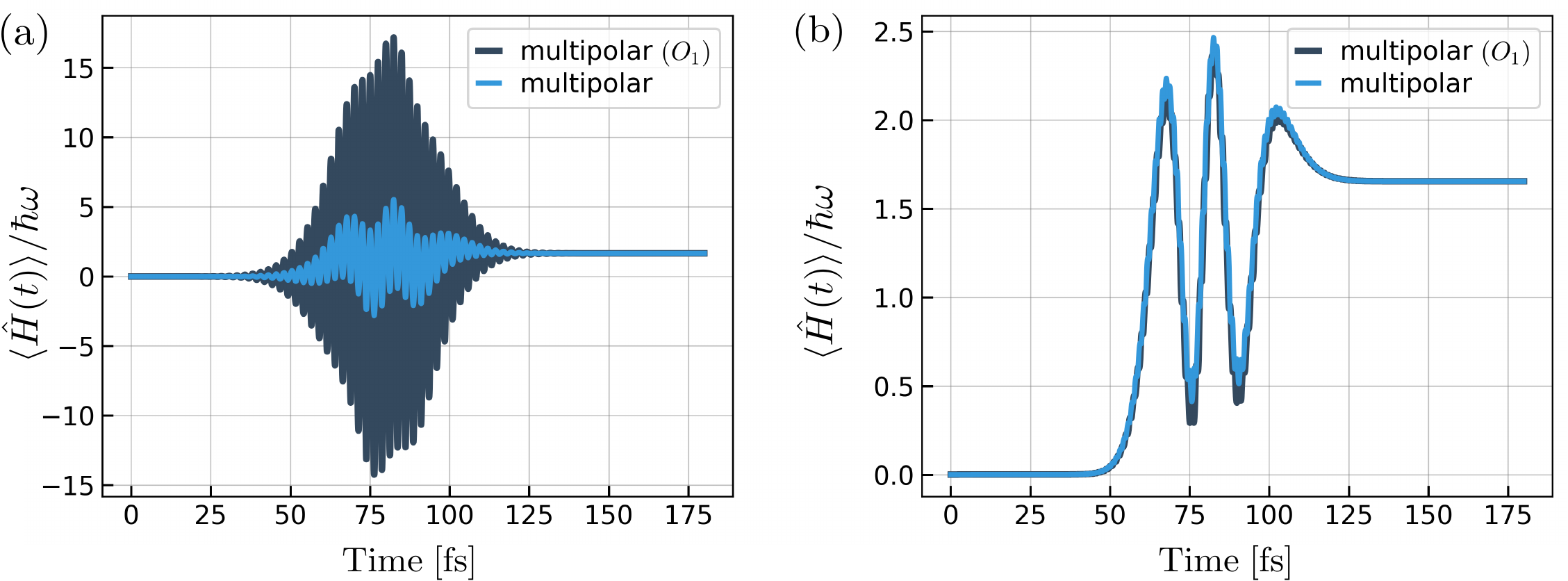}
\caption{The average energy of the chain with $N=80$ unit cells with two different origin of coordinates. The figure compares the dynamics generated by the multipolar PZW Hamiltonian (a) without and (b) with the charge density subtraction that guarantees charge neutrality in Eq. (21) from the main text.  Charge neutrality ensures that the light-matter interaction remains origin-independent. Simulation parameters are the same as in Figure 2 from the main text, except 0.2-$\mu$m-long chain.} 
\label{fig:origindependence}
\end{figure*}

\section{Convergence check for 2- and 6-MLWFs per unit-cell} \label{6vs2Appendix}

To demonstrate that the light-matter interactions beyond dipole approximation are converged with the number of bands,  in a manner akin to the  2 MLWFs per unit cell case in Sec. III A, we now parameterize the three highest valence and three lowest conduction bands of bulk \emph{t}PA and construct a tight-binding model with 6 MLWFs per unit cell with 6 nearest-neighbor couplings. We explicitly check the convergence of the dynamics considering nonuniform illumination of the chain as described in Sec. III D1.  We use Eq. (23) to compute the dynamics for 6-MLWFs case with $N_q=6$, and  Eq. (21) for the 2-MLWFs case.
Figure \ref{fig:6vs2orbital} shows that the dynamics for 2- and 6-MLWFs are indistinguishable demonstrating computationally favorable convergence with the number of bands. This is in contrast to velocity-gauge computations, where many bands are typically required for convergence.

\begin{figure*}[htb] 
\centering
\includegraphics[width=1.0\textwidth,keepaspectratio]{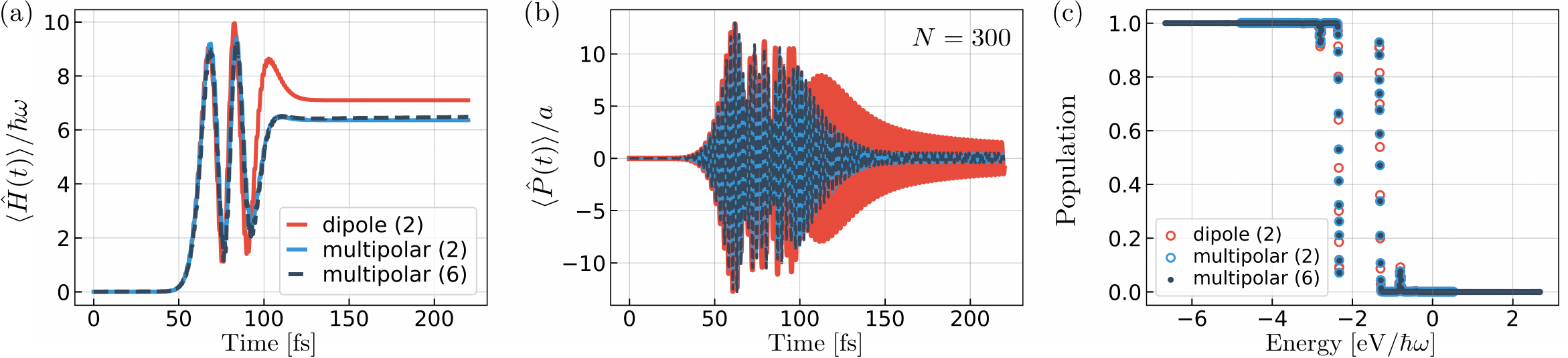}
\caption{The average energy (a), polarization (b) and population (c) of the  chain with $N=300$ unit cells comparing the dynamics generated by the multipolar PZW (blue) Hamiltonian [Eq. (21) from the main text] and dipole approximation (red) with 2-MLWFs per unit-cell, and the dynamics obtained with multipolar PZW (black) Hamiltonian [Eq. (23) from the main text] with 6-MLWFs per unit-cell. Simulation parameters are the same as in Figure 2 from the main text, except 0.75-$\mu$m-long chain.} 
\label{fig:6vs2orbital}
\end{figure*}

\section{Higher-order expansion terms for a Gaussian beam}    \label{appendix}

The general light-matter interaction term in the form of multipolar expansion in Eq. (6) from the main text, can be written compactly utilizing Einstein summation notation as
\begin{equation}
    \hat{H}_{\mathrm{LM}} = -\sum_j \sum_{n=1}^{\infty} \frac{1}{n!} q_j \ \hat{r}_j^{\nu_1} \dots \hat{r}_j^{\nu_n} \nabla_{\nu_2} \dots \nabla_{\nu_{n}} E_{\nu_1} (\mathbf{0},t),
\end{equation}
and the general $n$-th expansion term in terms of multipole operators $\hat{M}^{{\nu_1} \dots {\nu_n}} \equiv q_j \ \hat{r}_j^{\nu_1} \dots \hat{r}_j^{\nu_n} $ is given as 
\begin{equation}
    \hat{H}^{(n)} (t) = -\frac{1}{n!} \hat{M}^{{\nu_1} \dots {\nu_n}} \nabla_{\nu_2} \dots \nabla_{\nu_{n}} E_{\nu_1} (\mathbf{0},t).
 \end{equation}
Here $n=1,2,3,\dots$ and $n=1:$ dipole $\propto \hat{M}^{\nu_1} E_{\nu_1} $, $n=2:$ quadrupole $\propto \hat{M}^{\nu_1 \nu_2} \nabla_{\nu_2} E_{\nu_1} $, etc. For an $n$-th order multipolar term, the electric field is evaluated through its $(n-1)^{\mathrm{th}}$ spatial derivative at a chosen expansion point.   

In Section IV A from the main text, we consider a Gaussian beam [Eq. (26)] and to calculate the derivatives, we expand around the beam center $(x,y)=(0,0)$. Due to symmetry of the Gaussian beam around the beam center or origin, the odd-order derivatives are zero. Therefore, even-order multipolar terms like quadrupole $(n=2)$ or 16-pole $(n=4)$ do not contribute.  Note that for any other expansion point, the inversion symmetry will not be satisfied and the even-order multipolar contributions will not necessarily vanish.

We can directly calculate the octupole term $(n=3)$ at the origin as follows. 
\begin{equation} \label{OctupoleTerm}
    \hat{H}^{(3)}_{\mathrm{octupole}} (t) =  -\frac{1}{3!} \sum_j  q_j \ \hat{r}_j^{\nu_1}  \hat{r}_j^{\nu_2} \hat{r}_j^{\nu_3} \nabla_{\nu_2} \nabla_{\nu_{3}} E_{\nu_1} (\mathbf{0},t).
\end{equation}
As shown in Sec. III D1 from the main text, we can decompose the Gaussian beam as $\mathbf{E}(x,y,z=0,t)= E_x(x)E_y(y)E(t)\hat{e}_x$. There is no $z$-dependence in the plane $z=0$, so any derivative involving $\partial_z$ vanishes. Thus, Eq.~\ref{OctupoleTerm} reduces to
\begin{align} \label{octupole}
    \hat{H}^{(3)}_{\mathrm{octupole}} (t) &= -\frac{1}{3!} E(t) \sum_j  q_j (\hat{r}_j \cdot \hat{e}_x) \bigg[ E_y \hat{x}_j^2 \partial^2_x E_x  +  \hat{y}_j^2 E_x \partial^2_y E_y +  
     2 \hat{x}_j \hat{y}_j \partial_x E_x \partial_y E_y \bigg]_{(0,0)} \nonumber \\
    &= \frac{1}{3s^2} E(t) \sum_j  q_j \hat{x}_j \big[ \hat{x}_j^2 + \hat{y}_j^2 \big] \nonumber \\
    &= \frac{1}{3s^2} E(t) \sum_j  q_j \hat{x}_j\ \hat{r}_j^2,
\end{align}
where $ E_y(0)\partial^2_x E_x(0) = E_x(0) \partial^2_y E_y(0) = -2/s^2$ and $\partial_x E_x(0) = \partial_y E_y(0) =0$ due to symmetry around the beam center, and $\hat{r}_j^2 \equiv \hat{x}_j^2 + \hat{y}_j^2$. Similarly, the 32-pole $(n=5)$ correction term can be evaluated using the $4$th-order derivatives at the origin: $E_y(0)\partial^4_x E_x(0)  = E_x(0) \partial^4_y E_y(0) = 12/s^4$ and $\partial_x^2 E_x(0) = \partial_y^2 E_y(0) = 4/s^4$,
\begin{align} \label{32pole}
    \hat{H}^{(5)}_{\mathrm{32-pole}} (t) &= -\frac{1}{5!} E(t) \sum_j  q_j(\hat{r}_j \cdot \hat{e}_x) \bigg[E_y \hat{x}_j^4 \partial^4_x E_x  + \hat{y}_j^4 E_x \partial^4_y E_y +  
      6 \hat{x}_j^2 \hat{y}_j^2 \partial_x^2 E_x \partial_y^2 E_y \bigg]_{(0,0)} \nonumber \\
    &= -\frac{1}{10 s^4} E(t) \sum_j  q_j \hat{x}_j \big[ \hat{x}_j^4 + \hat{y}_j^4 + 2 \hat{x}_j^2 \hat{y}_j^2  \big] \nonumber \\
    &= -\frac{1}{10 s^4} E(t) \sum_j  q_j \hat{x}_j\ \hat{r}_j^4 .
\end{align}
Thus, the beyond-dipole multipolar terms for expanding the Gaussian beam at the beam center are given by:
\begin{equation}
    \hat{H}^{(2m+1)} (t) = \frac{(-1)^{m+1}}{(2m+1) m!} E(t) \frac{1}{s^{2m}} \sum_j q_j \hat{x}_j \hat{r}_j^{2m} .
\end{equation}
Note that the wavelength of light does not appear in these expressions and that when the spot size $s$ is much larger than the size of the material, higher-order terms become negligible relative to the dipole term. This result thus formally shows that under perpendicular illumination of a 1-D or 2-D material with a Gaussian beam, the long-wavelength limit is guaranteed to be satisfied. This is why in Sec. IV A from the main text, we isolate the effect of the nonuniform illumination alone. 

We can now write Eqs.~\ref{octupole} and ~\ref{32pole} in the basis of \emph{modified} MLWFs as done in Sec. IIIB from the main text. The octupole term is given as 
\begin{align} \label{octupoleExact}
    \hat{H}^{(3)}_{\mathrm{octupole}}(t) &= -|e| \frac{E(t)}{3 s^2} \sum_{\substack{\alpha_x,\beta_x \\ \mathbf{R}_1,\mathbf{R}'_1}} ( R'_{1x} + {x}_{\beta} ) \bigg( ( R'_{1x} + {x}_{\beta} )^2 + \sum_{\alpha_y \mathbf{R}_2} \langle \beta_x \mathbf{R}'_1 | \alpha_y \mathbf{R}_2 \rangle (R_{2y} + {y}_{\alpha})^2 \langle \alpha_y \mathbf{R}_2 | \alpha_x \mathbf{R}_1 \rangle  \bigg) \nonumber \\
    & \times  \bigg[ \hat{c}^{\dagger} _{\beta_x \mathbf{R}'_1} \hat{c}_{\alpha_x \mathbf{R}_1}  -  \rho_{\beta_x \mathbf{R}'_1,\alpha_x \mathbf{R}_1}(0) \delta_{\beta_x,\alpha_x} \delta_{ \mathbf{R}'_1, \mathbf{R}_1} \bigg] ,
\end{align}
and 32-pole term is given as
\begin{align} \label{32poleExact}
    \hat{H}^{(5)}_{\mathrm{32-pole}}(t) &= |e| \frac{E(t)}{10 s^4}  \sum_{\substack{\alpha_x,\beta_x \\ \mathbf{R}_1,\mathbf{R}'_1}} ( R'_{1x} + {x}_{\beta} ) \bigg( ( R'_{1x} + {x}_{\beta} )^4 + \sum_{\alpha_y \mathbf{R}_2} \langle \beta_x \mathbf{R}'_1 | \alpha_y \mathbf{R}_2 \rangle (R_{2y} + {y}_{\alpha})^4 \langle \alpha_y \mathbf{R}_2 | \alpha_x \mathbf{R}_1 \rangle +  \nonumber \\
    & 2 ( R'_{1x} + {x}_{\beta} )^2 \sum_{\alpha_y \mathbf{R}_2} \langle \beta_x \mathbf{R}'_1 | \alpha_y \mathbf{R}_2 \rangle (R_{2y} + {y}_{\alpha})^2 \langle \alpha_y \mathbf{R}_2 | \alpha_x \mathbf{R}_1 \rangle \bigg) \bigg[ \hat{c}^{\dagger} _{\beta_x \mathbf{R}'_1} \hat{c}_{\alpha_x \mathbf{R}_1}  - \nonumber \\
    &   \rho_{\beta_x \mathbf{R}'_1,\alpha_x \mathbf{R}_1}(0) \delta_{\beta_x,\alpha_x} \delta_{ \mathbf{R}'_1, \mathbf{R}_1} \bigg] .
\end{align}
We use the Eqs.~\eqref{octupoleExact} and ~\eqref{32poleExact} to calculate corrections to the dipole term [Eq. (21)] upto octupole and 32-pole term, respectively in Sec. IV A. 

\section{Computational cost comparison} 

\begin{figure*}[htb] 
\centering
\includegraphics[width=1.0\textwidth,keepaspectratio]{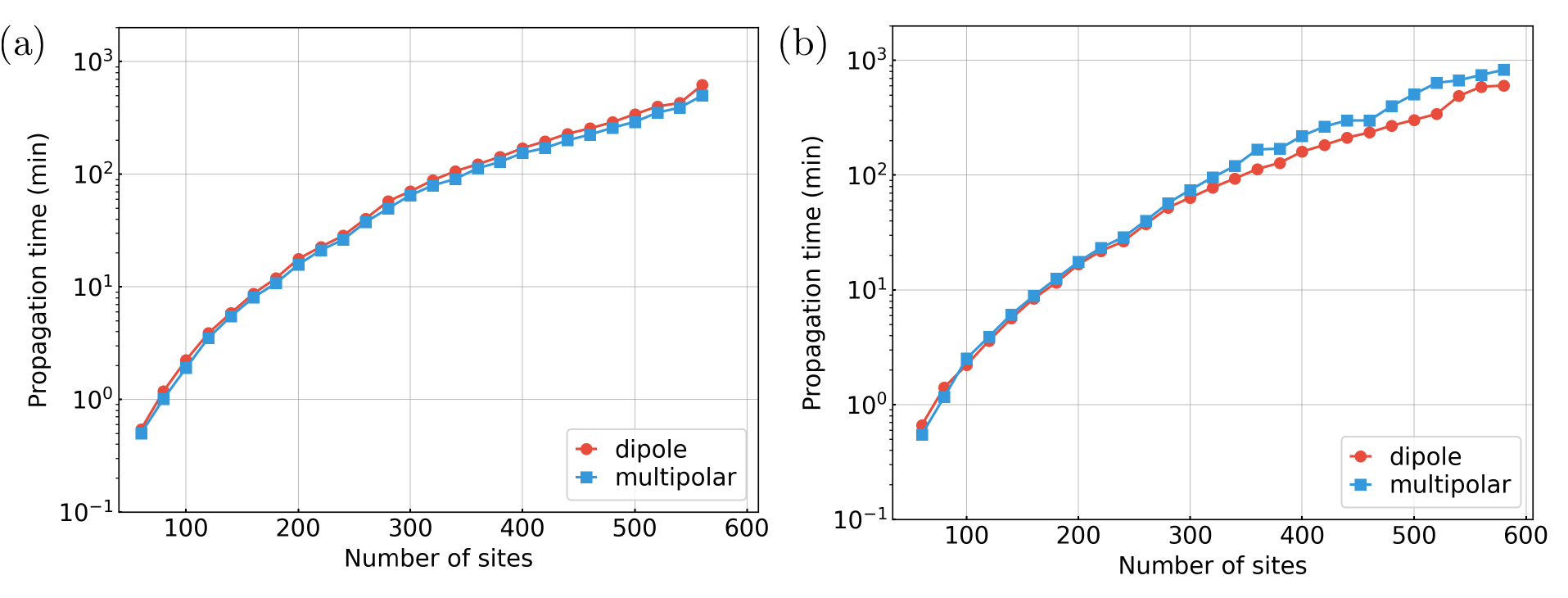}
\caption{Comparison of the computational cost on a log scale of the dipole versus the multipolar when the electric field is (a) separable into spatial and temporal components, and (b) non-separable. All parameters are the same as in Fig. 3(a) and Fig. 6(a) from the main text, respectively.    }
\label{fig:computational}
\end{figure*}

As discussed in Sec.~IV E of the main text, if the electric field is separable into spatial and temporal components. Thus, the temporal part in Eqs. 21 and 23 from the main text can be factorized out. Here, we compare computational cost (on a log scale) to reproduce Fig. 3(a) from the main text for the dipole vs the multipolar case. Figure~\ref{fig:computational}(a) shows the computational cost for the dipole and multipolar is almost identical in this case. In the second case, we compare computational cost to reproduce Fig. 6(a) from the main text, where electric field is not separable into spatial and temporal components. In Fig.~\ref{fig:computational}(b), we observe that computational cost of multipolar moderately increases in comparison to the dipole as the system size increases but remains of the same order. Importantly, both methods exhibit nearly identical scaling with system size, indicating that the multipolar formulation increases only the computational prefactor rather than the overall scaling.  

\end{widetext}

\end{document}